# Anisotropic implantation damage build-up and crystal recovery in $\beta$-Ga$_2$O$_3$

D. M. Esteves[1,2,*], S. Magalhães[2,3], Â. R. G. da Costa[4], K. Lorenz[1,2,3], M. Peres[1,2,3]

[1]     INESC Microsystems and Nanotechnology, Rua Alves Redol 9, Lisboa 1000-029, Portugal

[2]     Institute for Plasmas and Nuclear Fusion, Instituto Superior Técnico, University of Lisbon, Av. Rovisco Pais 1, Lisboa 1049-001, Portugal

[3]     Department of Nuclear Science and Engineering, Instituto Superior Técnico, University of Lisbon, Estrada Nacional 10, km 139.7, Bobadela 2695-066, Portugal

[4]     Centro de Ciências e Tecnologias Nucleares, University of Lisbon, Estrada Nacional 10, km 139.7, Bobadela 2695-066, Portugal

*     Corresponding author: duarte.esteves@tecnico.ulisboa.pt

**ABSTRACT:**

The present work aims at investigating the defect accumulation and recovery dynamics in the inherently anisotropic $\beta$-Ga$_2$O$_3$ lattice. A systematic Rutherford Backscattering Spectrometry in Channelling mode (RBS/C) analysis of Cr-implanted samples was performed across multiple surface orientations and channelling directions. Distinct apparent defect accumulation and annealing rates were observed along different channelling axes, mainly attributed to the shadowing of certain types of defects along some directions. The efficient defect removal observed after annealing was correlated with the strain relaxation observed via High-Resolution X-ray diffraction (HRXRD) at temperatures as low as 500 °C, which is attributed to the removal of point defects. Annealing at higher temperature further improves crystalline quality but at a slower rate. In short, this work enhances the understanding of the effect of structural anisotropic properties of $\beta$-Ga$_2$O$_3$ during ion implantation, as well as the crystal recovery during thermal annealing, highlighting the interplay between crystallography and defect dynamics.

1. Introduction

The future of electronics depends on finding novel materials with properties that allow them to meet ever-more-stringent performance requirements. In this context, the oxide wide bandgap semiconductor β-Ga$_2$O$_3$ is one of the most prominent candidates for future high-power electronics [1] and (opto)electronic devices [2], mainly due to its wide bandgap of ~4.9 eV at room temperature (RT) and its large breakdown electric field of ~8 MV/cm [3], which is greater than those of GaN or SiC. On the other hand, the wide bandgap endows it with a high transparency [4] down to ~250 nm, enabling its usage in optoelectronic applications such as solar-blind photodetectors [5–7].

Nowadays, ion implantation is a routine technique used in the Si industry, with over 50 years of history [8]. However, the necessary fundamental studies required to enable the standardised application of this technique to wide bandgap semiconductors, in particular β-Ga$_2$O$_3$, are still underway. In fact, in recent years, multiple ion implantation and irradiation studies have shown intriguing effects, such as a disorder-induced-ordering phenomenon triggering a phase transformation to the γ-phase [9]. On the other hand, as a monoclinic material, the elastic anisotropy poses additional challenges, but also offers additional tuning parameters that can be exploited in innovative ways for strain or defect engineering [9,10]. Conversely, the physical mechanisms governing the defect creation and accumulation dynamics under ion implantation in this material, as well as its removal under thermal annealing, are complex and are not yet fully understood.

While there are several recent works addressing the study of the implantation effects in β-Ga$_2$O$_3$ by Rutherford Backscattering Spectrometry in Channelling mode (RBS/C) [9–18], only a few of them focus specifically on assessing the anisotropy of the monoclinic system by comparing implantation into bulk crystals with different surface orientations [19–21]. Moreover, to the best of our knowledge, no systematic study has yet been performed regarding the RBS/C signal obtained by channelling along different crystallographic axes for a given surface orientation. Such a study is important in order to understand the effect of surface orientation on implantation damage build-up. Additionally, taking the particularities of the base-centred monoclinic structure into account, the clear identification and assessment of multiple axes also contributes to the establishment of reproducible and standardised methodologies to assess important properties of this material, such as its crystalline quality.

In order to address this challenge, in this work, we perform a systematic experimental RBS/C study of the defect accumulation in β-Ga$_2$O$_3$ samples with different surface orientations, as well as of the lattice recovery under annealing. These studies are complemented by a strain analysis via High-Resolution X-ray Diffraction (HRXRD), with both techniques showing a good agreement. Moreover, focusing on the ($\bar{2}$01)-oriented samples, we perform a systematic RBS/C study on several axes, considering both increasing implantation fluences and annealing temperature.

## 2. Materials and methods

The samples used in this work were unintentionally-doped commercial $\beta$-Ga$_2$O$_3$ single-crystals from Novel Crystal Technology. These were grown by the edge-defined film-fed growth (EFG) method and cut along the (100), (010), (001) and $(\bar{2}01)$ planes, yielding 5 mm × 5 mm samples with a thickness of ~500 μm.

The 250 keV Cr$^+$ implantations were performed simultaneously for all surface orientations, at room temperature, using the 500 kV implanter of the Ion Beam Centre of the Helmholtz Zentrum Dresden-Rossendorf (IBC-HZDR), to fluences up to $2\times10^{14}$ cm$^{-2}$, with a tilt angle of 7° to avoid channelling. The annealing of the samples was performed in an N$_2$ atmosphere using a Rapid Thermal Processor AS-One 100 by AnnealSys.

The RBS/C measurements were performed using the 2.5 MV Van de Graaff accelerator of the Laboratory of Accelerators of Instituto Superior Técnico (IST), Universidade de Lisboa [22]. A 2 MeV He$^+$ ion beam with a nominal current of ~4 nA was employed, with a Si PIN diode detector placed at an angle of 140° with respect to the beam direction. The random spectra were obtained by tilting the sample away from the channelling condition and rotating the azimuthal angle. The relative defect concentration profiles were extracted by applying the two-beam approximation, using the DECO software developed at the University of Jena [23,24]. This model considers the analysing beam as being composed of two contributions: the random fraction, which interacts with all the lattice atoms as if the target was amorphous, and the aligned fraction, which interacts only with lattice atoms that are displaced from their equilibrium positions, regardless of whether it is due to the presence of defects or due to thermal oscillations [23–25]. The model allows one to analytically calculate the probability of an ion transitioning from the aligned to the random fraction due to de-channelling [26]. However, the model does not properly account for more complex extended defects, which also contribute to de-channelling [27,28]. In this paper, the critical angle of channelling (i.e., the minimum angle for which the impinging particle leaves the channel) was used to compensate for this shortcoming of the model, i.e., the effective critical angle became a phenomenological parameter reflecting additional de-channelling mechanisms. The results of this type of procedure have been previously compared with state-of-the-art RBS/C analysis using Monte Carlo codes in the case of GaN, showing a good agreement [29]. The critical angle pre-factors used in the present work varied between 0.10 and 0.40, likely reflecting the contribution of extended defects and lattice distortions that are not fully captured within the analytical framework of the two-beam approximation.

While the literature is quite scarce regarding the application of the two-beam model to $\beta$-Ga$_2$O$_3$ [10–12], the defect profiles extracted in the present work are consistent and in good agreement with Stopping and Range of Ions in Matter (SRIM) Monte Carlo vacancy distributions [30]. These simulations were performed in the full damage cascades calculation mode, using displacement energies of 28 eV for Ga and 14 eV for O [31] and a density of 5.88 g/cm$^3$ for $\beta$-Ga$_2$O$_3$ [32]. Based on the original Fortran code and the Götz and Gärtner parametrisation for the universal de-channelling function [26], a Python script was developed in the context of this work to apply the two-beam approximation, and is freely available on GitHub: https://github.com/DuarteME/pyDECO

The HRXRD measurements were performed at the Bruker D8 Discover diffractometer of the Laboratory of Accelerators of IST. The primary beam optics consists of a Göbel mirror, a 0.2 mm collimation slit, and a 2-bounce (220)-Ge monochromator, to select the copper (Cu) K$\alpha_1$ X-ray line (wavelength of 1.5406 Å). The secondary beam path entails a 0.1 mm slit and a scintillation detector.

3. Results and discussion

3.1. RBS/C in monoclinic $\beta$-Ga$_2$O$_3$

Fig. 1 shows the conventional unit cell of $\beta$-Ga$_2$O$_3$, along with the corresponding lattice parameters $a$, $b$, $c$ and $\beta$. Additionally, the most common surface planes have been marked, namely the (100), (010), (001) and $(\bar{2}01)$ planes, and the corresponding normal vectors have been indicated. This figure thus shows how different surface orientations and plane normals are related to one another in this monoclinic system.

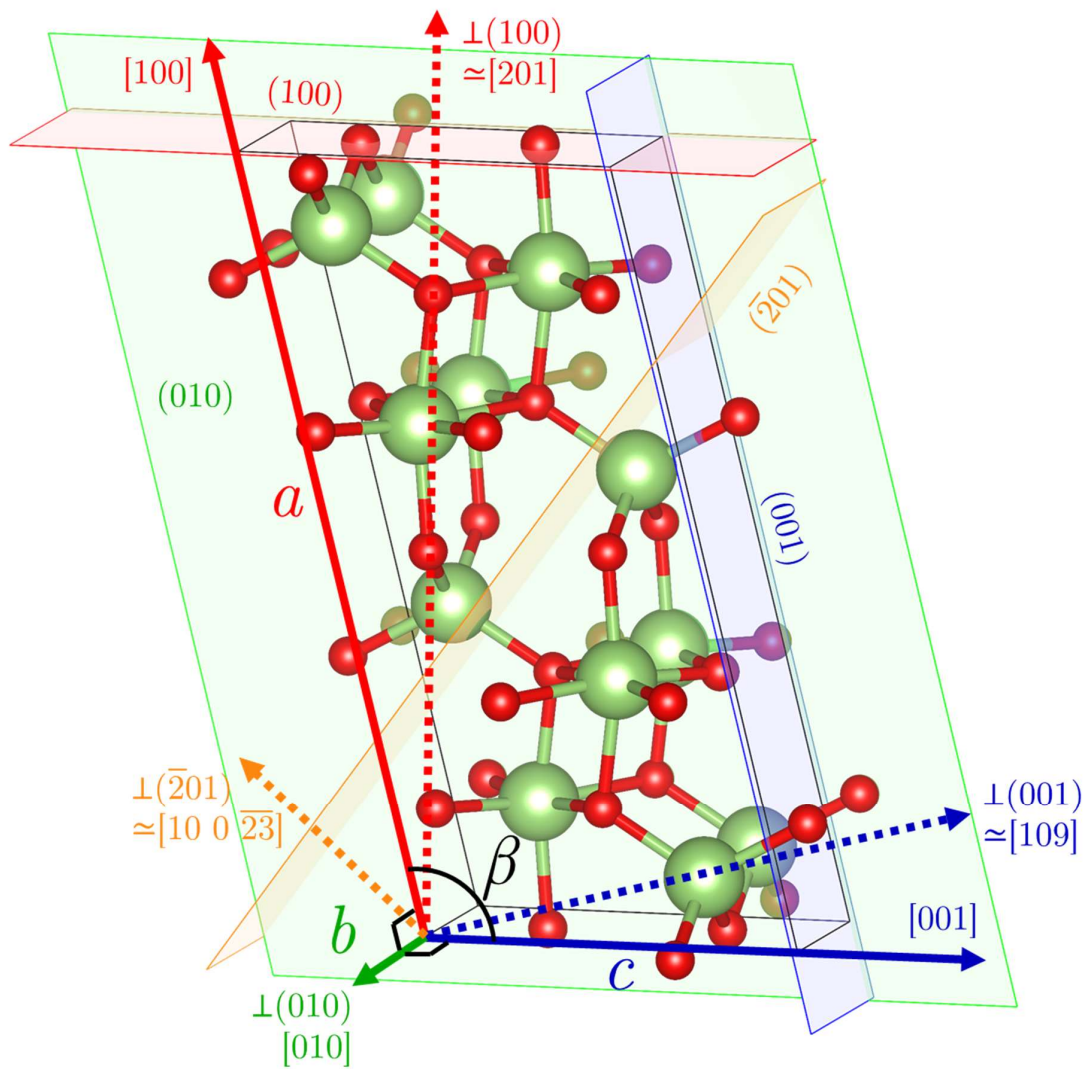

**Fig. 1** | Conventional unit cell of $\beta$-Ga$_2$O$_3$, where the Ga atoms are shown in green and the O atoms are shown in red. The red, green, blue and orange planes correspond, respectively, to the (100), (010), (001) and $(\bar{2}01)$ planes, and the dashed vectors of the same colour show the directions perpendicular to the respective plane. The figure was obtained using the Vesta software [33].

For the (010)-oriented sample, the normal vector corresponds exactly to [010], since $\boldsymbol{b}$ is the unique axis of the monoclinic system. However, the angle $\beta$ between $\boldsymbol{a}$ and $\boldsymbol{c}$ is not 90°, which implies that, for (100)-oriented samples, the surface normal is parallel to the vector $c \csc\beta\, \boldsymbol{a} - a \cot\beta\, \boldsymbol{c}$, while, for (001)-oriented samples, the surface normal is parallel to the vector $-c \cot\beta\, \boldsymbol{a} + a \csc\beta\, \boldsymbol{c}$. Due to the presence of the $\beta$-dependent terms, it is not possible to express these plane normals as linear combination of the basis vectors with integer coefficients. Since $-\frac{c\,\mathrm{cs}}{a \cot\beta} \simeq 2$ and $-\frac{a \csc\beta}{c \cot\beta} \simeq 9$, we consider [201] and [109] as the axes closest to the normal of the (100) and (001) planes.

In order to perform these calculations, we employed a Mathematica script, freely available online [34], as well as the lattice constants (obtained by X-ray diffraction): $a = 12.2628$ Å, $b = 3.0446$ Å, $c = 5.8235$ Å and $\beta = 103.807°$. Specifically, for the monoclinic system, the angle between the direction [$uvw$] and the normal to the plane ($hkl$) can be written as:

$$\theta = \arccos\left[\frac{hu + kv + lw}{\sqrt{a^2u^2 + b^2v^2 + c^2w^2 + 2acuw \cos\beta}\sqrt{\left(\frac{h^2}{a^2} + \frac{k^2}{b^2} + \frac{l^2}{c^2}\right)\csc^2\beta - \frac{k^2}{b^2}\cot^2\beta - \frac{2hl}{ac}\cot\beta \csc\beta}}\right], \quad (1)$$

which yields angles of about 0.1° and 0.3° between [201] and the normal to the (100) plane, and between [109] and the normal to the (001) plane, respectively. For the $(\bar{2}01)$ plane, the normal is parallel to the vector $\boldsymbol{a} - \frac{a\,(a+2c\cos\beta)}{c\,(2c+a\cos\beta)}\boldsymbol{c}$. Since $-\frac{a\,(a+2c\cos\beta)}{c\,(2c+a\cos\beta)} \simeq -2.3$, we will consider $[10\,0\,\overline{23}]$ as the vector closest to the normal to this plane; indeed, Eq. (1) shows that the angle between $[10\,0\,\overline{23}]$ and the normal to $(\bar{2}01)$ is about 0.1°.

As an example, this work contains a detailed study of different channelling axes in $(\bar{2}01)$-oriented samples. The samples were pre-aligned via X-ray diffraction in such a way that the [010] direction is vertical, so that the probed axes for a horizontal RBS/C scan (in $\vartheta$) all belong to the (010) trace, i.e., they are of the form [$u0w$]. Fig. 2 shows selected portions of an angular $\vartheta$-scan performed from –38 to +42° in a pristine sample, along with the identification of the main axes. It is noteworthy that the "normal" channel $[\overline{10}\,0\,23]$ does not exhibit a very good minimum yield in comparison with the level of planar channelling. Since there are two close-by channels (within ~3° of the plane normal), namely $[\bar{2}05]$ and $[\bar{1}02]$, with much better channelling qualities, care should be taken when reporting RBS/C results for this surface, as this may lead to an incorrect interpretation of the data. As $[\bar{2}05]$ is closer to the normal and has the better minimum yield, this channel was selected to be explored in the current work. Moreover, there are two additional channels with very good minimum yields at larger angles, namely $[\bar{1}00]$ and [001], which are parallel to the $\boldsymbol{a}$ and $\boldsymbol{c}$ lattice vectors, respectively. Tab. I summarises the tilt angles between the surface normal and each of these channels, as given by Eq. (1). The angular positions of these axes are also shown in the stereographic projection in the lower panel of Fig. 2.

| Axis | [001] | $[\bar{2}05]$ | $[\overline{10}\,0\,23]$ | $[\bar{1}02]$ | $[\bar{1}00]$ |
|---|---|---|---|---|---|
| $\theta$ (°) | –36.215 | –1.958 | 0.098 | 3.038 | 39.78 |

**Tab. I** | Tilt angle between the $(\bar{2}01)$ surface normal and each of the directions observed in Fig. 2.

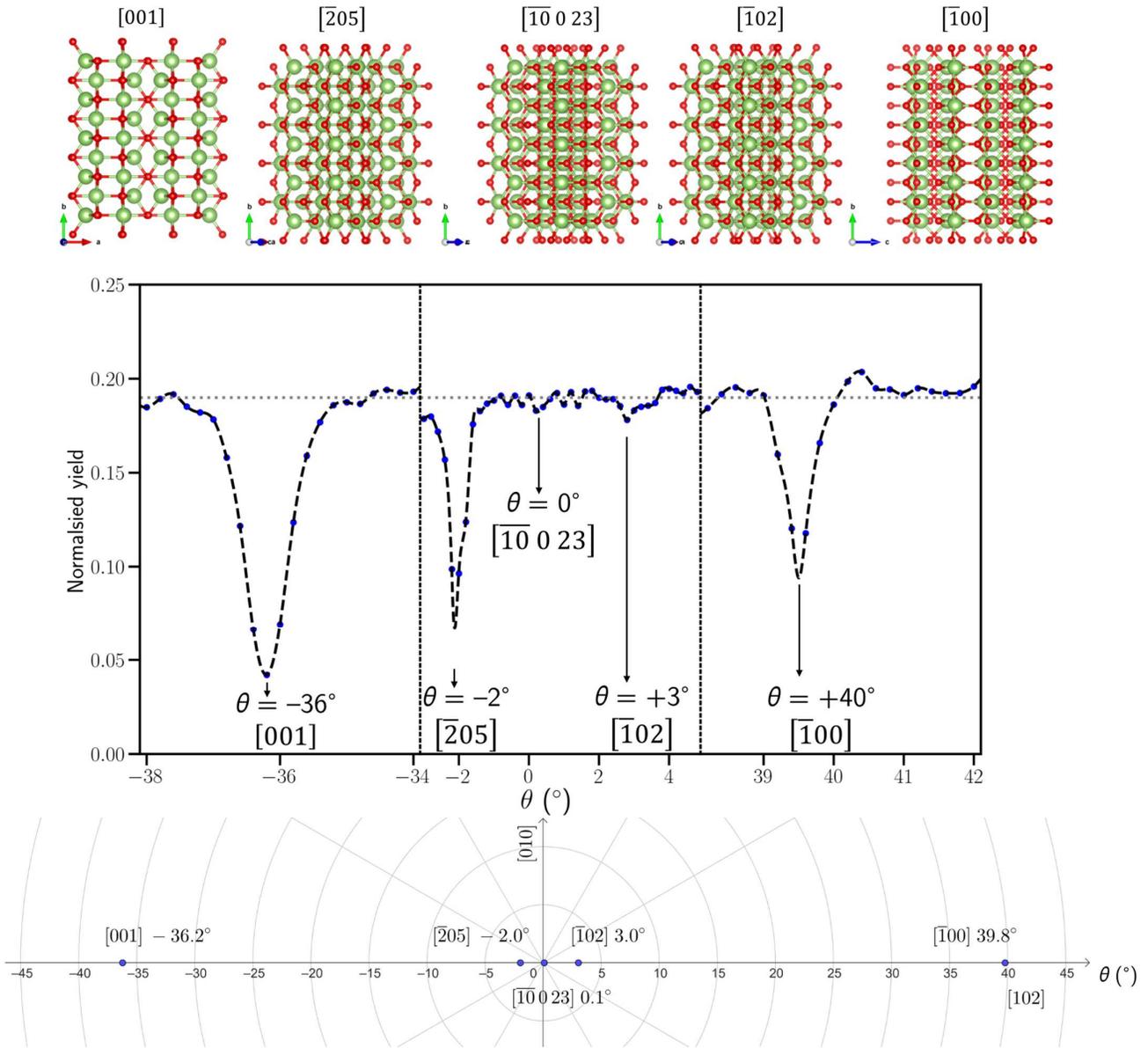

**Fig. 2** | RBS/C horizontal angular scan for $(\bar{2}01)$-oriented sample with the **b** axis in the vertical direction. The sign of the $\vartheta$ angle was chosen in such a way that the rotation direction is the one induced by **b**, according to the right-hand rule. The blue dots are experimental data points, while the dashed lines are guides to the eye. The dotted line indicates the level of planar channelling and the yield is normalised to the random level. The lower panel shows a stereographic projection of the axes within the (010) plane. The figures above show sketches of the view of the crystal lattice along the aligned axes, obtained using the VESTA software [33].

In order to better assess the channelling quality associated with each of the found crystallographic axes, an RBS/C study was performed on the pristine, as-grown samples cut with different surface orientations. As per the discussion above, the sample with the $(\bar{2}01)$ surface orientation was measured along different axes. Fig.

3 shows the RBS/C spectra of the pristine samples after alignment, as well as a random spectrum for reference. The Ga and O barriers are marked with arrows.

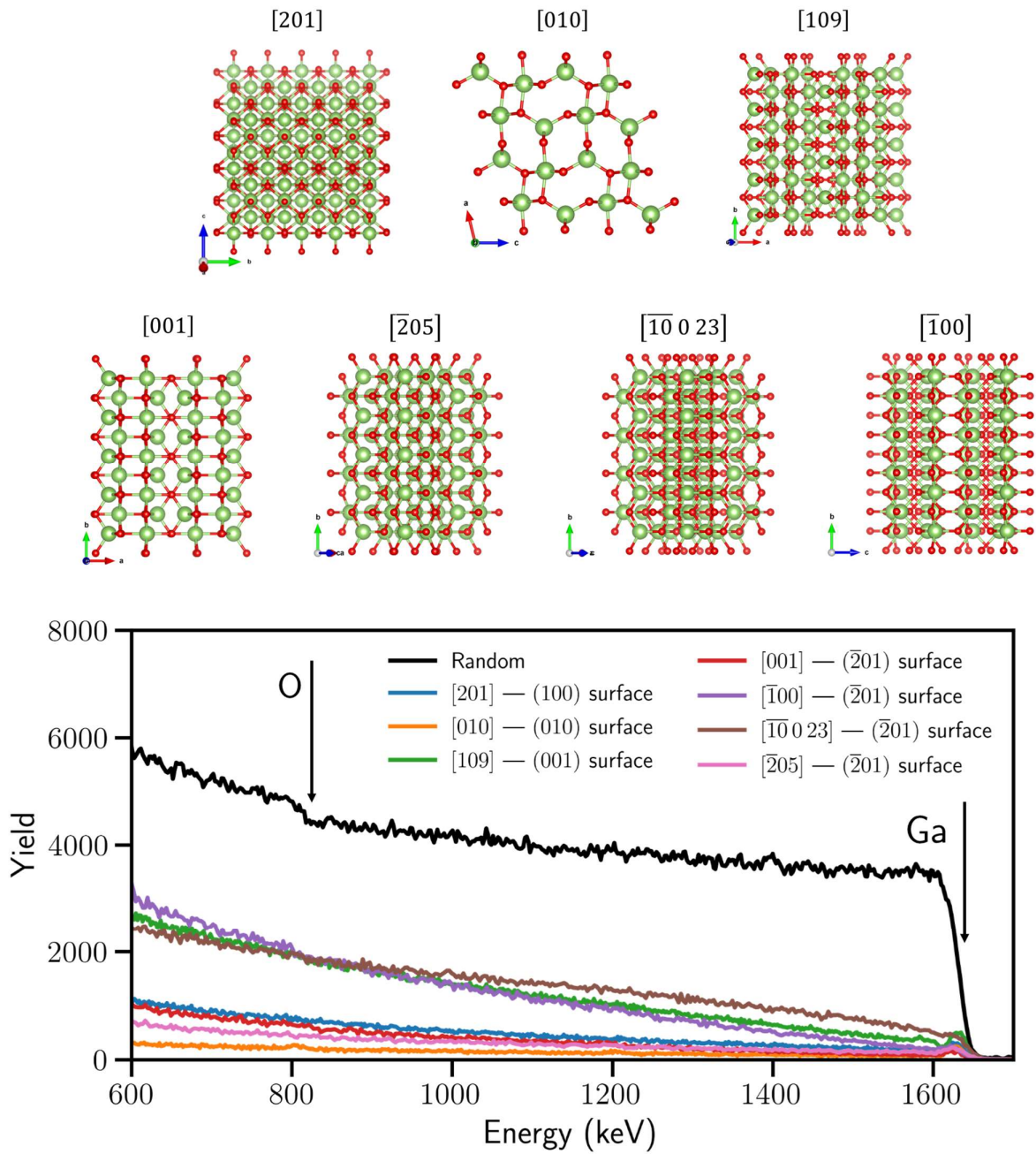

**Fig. 3** | RBS/C spectra of the pristine samples with different surface orientations and along different crystallographic axes, as well as a random spectrum for reference. The figures above show sketches of the view of the crystal lattice along the aligned axes, obtained using the VESTA software [33].

Fig. 3 reveals that the channelling quality is different along each crystallographic axis. In fact, it is possible to distinguish some directions along which there is an enhanced de-channelling, namely $[\bar{1}00]$, $[109]$ and $[\bar{10}\,0\,23]$, from others where de-channelling is not so prevalent, namely $[010]$, $[\bar{2}05]$, $[001]$ and $[201]$. With the exception of the $[\bar{10}\,0\,23]$ channel, all the spectra show a clear peak at about 1625 keV, which is related with direct backscattering from the surface. Regarding the [109], the spectral shape is intriguing, and there is already some structure present in the pristine sample, consisting of oscillations likely related with a resonant de-channelling effect, which has been observed before in the transition from axial to planar channelling [35,36]. Similar de-channelling behaviours with respect to the channels perpendicular to different surface orientations have been previously reported in the literature by J. Matulewicz et al. [19].

By comparing the yield of the aligned spectrum in a given energy window, $Y_{aligned}$, with that of the random spectrum $Y_{random}$, it is possible to calculate the minimum yield associated with each axis, given by $Y_{aligned}/Y_{random}$, which provides information regarding crystalline or channelling quality. In the present case, we are probing the channelling quality, as the used samples were obtained by cutting bulk crystals along different planes. Tab. II shows that the minimum yield of the pristine samples in the 1500–1600 keV range can be as low as 2% when evaluated along the [010] channel and increases to 3–4% when evaluated along the [001] and [$\bar{2}$05] channels. These small values confirm the high single-crystalline quality of the samples. However, the minimum yield increases already to 6–8% when evaluated along the [201] and [$\bar{1}$00] channels, and can be as high as 12–19% if evaluated along the [109] and [$\bar{1}$0 0 23] channels. These differences in minimum yield are also consistent with the perceived channel quality suggested by the sketches in Fig. 3. In short, this study highlights the importance of stating the channelling axis when reporting RBS/C results, especially when performed with the goal of assessing crystalline quality.

| Channel | Minimum yield (%) |
| --- | --- |
| [010] | 2.1 |
| [001]* | 2.8 |
| [$\bar{2}$05]* | 4.2 |
| [201] | 5.6 |
| [$\bar{1}$00]* | 7.8 |
| [109] | 11.7 |
| [$\bar{1}$0 0 23]* | 18.6 |

**Tab. II** | Minimum yields calculated for the pristine samples from the data in Fig. 3, considering the energy range between 1500 and 1600 keV. The axes are ordered from the best (lowest minimum yield) to the worst (highest minimum yield) channelling quality. The channels measured within the same sample were marked with *.

3.2. Assessing implantation damage accumulation via RBS/C

In order to compare the evolution of the defect profiles with implantation fluence, a set of samples were implanted with 250 keV Cr$^+$ ions up to a fluence of $2\times10^{14}$ cm$^{-2}$. For each implantation fluence, the samples were measured by RBS/C along the axis closest to their surface normals, and the corresponding aligned spectra are shown in Fig. 4, as well as the corresponding random spectrum for reference. For each set of experimental data, a relative defect concentration profile was calculated using the DECO software [23,24].

Along with the aforementioned well-pronounced surface peak, the spectra of the implanted samples show an increased backscattering yield in the 1500–1600 keV region, associated with the implantation defects. The peak is due to direct backscattering at randomly displaced atoms that can occur mainly due to point defects, small defect clusters or amorphous regions. On the other hand, extended defects bend the channels and promote de-channelling, while leading to little direct backscattering [27]. In this case, the particles leave the channel and will be backscattered deeper in the sample, thus contributing to the lower-energy part of the spectrum.

As the implantation fluence increases, the yield of the direct backscattering peak increases steadily for the shown channels, albeit at different rates. It is important to note that, in agreement with previous studies [10], the surface layer of the (100)-oriented sample was detached and curled up at a fluence of $1 \times 10^{14}$ cm$^{-2}$, and thus the implanted layer was removed. For this reason, only the two lowest fluences are shown in Fig. 4 for this orientation. While for the (010)- and ($\bar{2}$01)-oriented sample, the aligned spectra never reach the random level, the aligned spectrum for the highest fluence obtained for the (001)-oriented samples is partially superimposed with the random spectrum, which may suggest amorphization. Indeed, the presence of an amorphous layer at the surface has been observed before in the (010) and ($\bar{2}$01) orientations via RBS/C, and confirmed by transmission electron microscopy (TEM) and selected area electron diffraction (SAED), although its onset occurs at a larger displacements per atom (dpa) level at the maximum of the defect profile (~11 dpa [20]). It should be pointed out that the $\beta$-to-$\gamma$ phase transformation typically takes place at about ~0.78 dpa [9,37,38], which is just above the maximum dpa level employed here (~0.72 dpa). However, no clear indications of the presence of the $\gamma$-phase were found in the present samples, at least considering the XRD and RBS/C sensitivities; the presence of localised amorphous or $\gamma$-phase clusters cannot be excluded, as these have been previously reported as common point-like defects in this material [39,40]. Moreover, considering the crystallographic relation between the two polymorphs, there is no high-symmetry axis of the $\gamma$-phase, along the normal to the (001) plane of the $\beta$-phase, which can also contribute to the enhanced backscattering yield measured here.

The concentration profiles obtained via DECO shows Gaussian-like depth distributions that extend approximately up to 200 nm in depth for all the surface orientations. Considering the ($\bar{2}$01), (010) and (001) surface orientations, the maximum of the relative defect concentration increases steadily up to about 0.4, 0.6 and 1.0, respectively, for the last fluence. Moreover, these profiles are in excellent agreement with the SRIM vacancy profile both regarding the damaged depth and the general location of the maximum. On the other hand, there is also an increased defect density at the surface, which has been attributed to the fact that the surface can act as a sink for diffusing defects [12]. With the possible exception of the (001)-oriented samples, no amorphization was detected in this study.

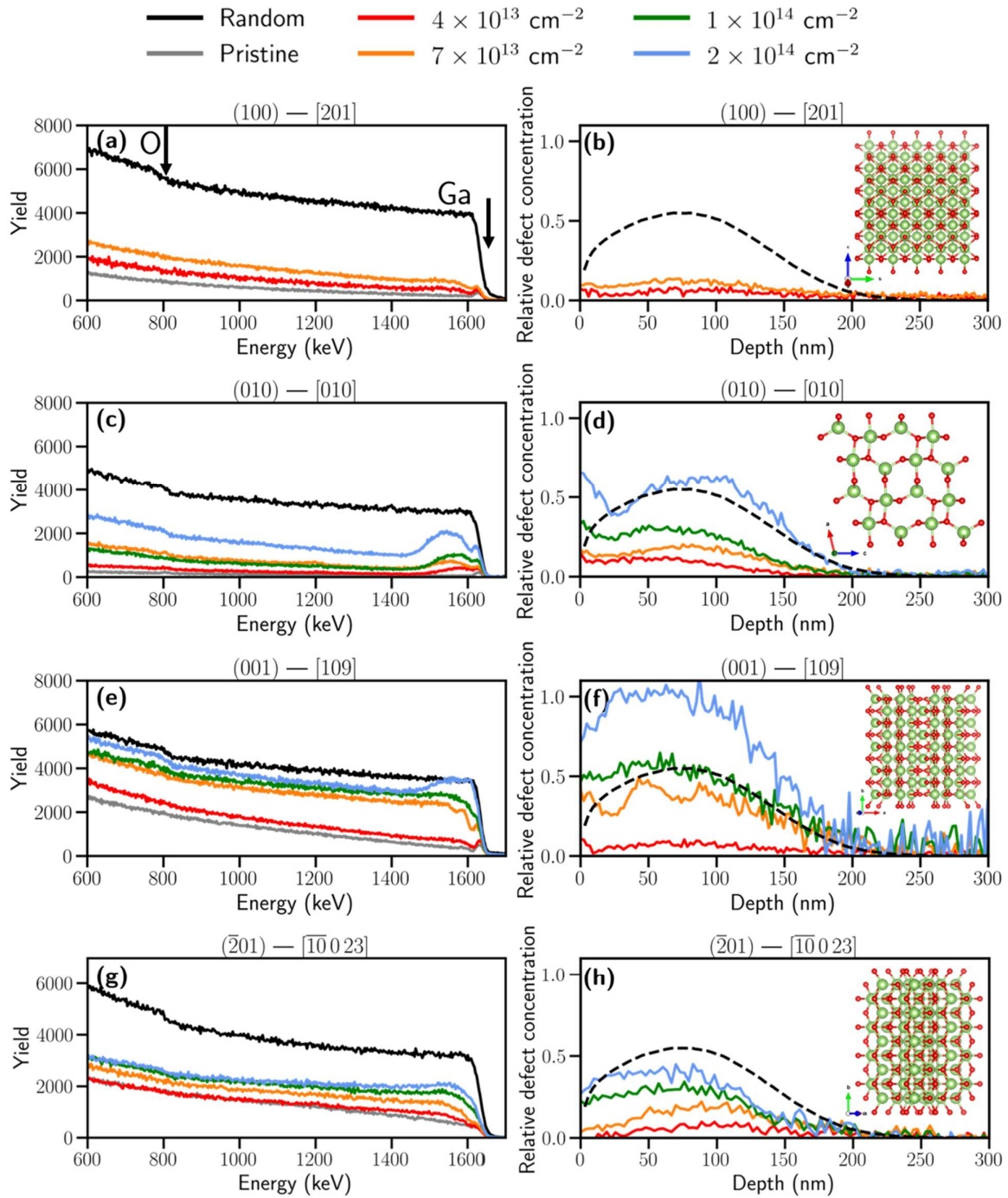

**Fig. 4** | RBS/C analysis for the (100)-oriented samples (a–b), along the [201] channel; (010)-oriented samples (c–d), along the [010] channel; (001)-oriented samples (e–f), along the [109] channel; and ($\bar{2}$01)-oriented samples (g–h), along the [$\overline{10}$ 0 23] channel, for different implantation fluences. Panels (a), (c), (e) and (g) show the experimental spectra, while panels (b), (d), (f) and (h) show the relative defect concentration profiles as calculated by DECO. The dashed lines represent the vacancy profiles obtained from SRIM Monte Carlo simulations, in an arbitrary scale. The insets show a sketch of the view of the crystal structure along the aligned axes, obtained using the VESTA software [33].

In order to better assess the damage accumulation with fluence, Fig. 5 shows the evolution of the quantity $\Delta\chi_{min} = (Y_{aligned} - Y_{pristine})/Y_{random}$ with the fluence, where $Y_{pristine}$ refers to the yield of the pristine sample in the aligned geometry. This quantity is similar to the minimum yield employed above, but allows one to probe only the implantation effects by discounting the component of the spectrum already present in the pristine samples [26]. The effect of annealing also shown in this figure will be discussed below.

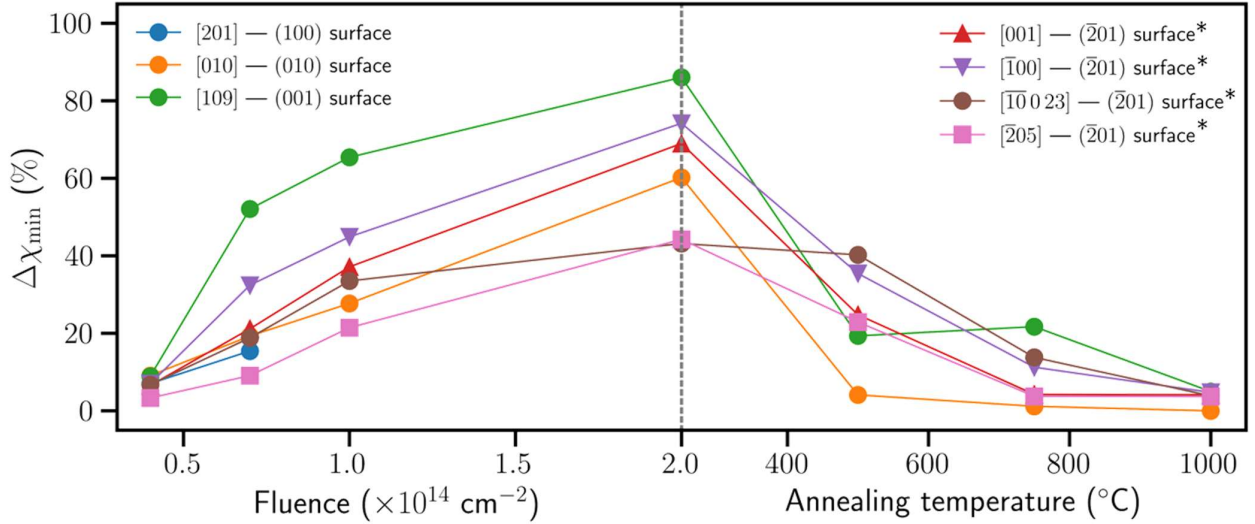

**Fig. 5** | Evolution of $\Delta\chi_{min}$ with the implantation fluence (left-hand side) and annealing temperature (right-hand side) for the different probed channels, as calculated from the spectra shown in Figs. 4, 6, 7 and 8. The dashed line separates the damage accumulation study from the thermal annealing study. The circle symbols refer to the channels perpendicular to the indicated surface planes. Channels marked with * belong to the same sample.

The evolution of $\Delta\chi_{min}$ with the fluence along different channels is quite interesting. In fact, as expected from the discussion above, the damage accumulation seems to be highest along the [109] direction. On first sight, one may attribute this to the reduced channelling quality. However, the initial channelling quality of the $[\overline{10}\,0\,23]$ axis was even worse, while the damage accumulation does not seem to have a very large impact and essentially saturates at about 45% after the $1\times10^{14}$ cm$^{-2}$ fluence. Along the [010] axis, the defect accumulation appears to be almost linear with the fluence, achieving a moderate value of about 60%. Furthermore, looking at the spectra in Fig. 4 (c), this direction reveals the most pronounced direct backscattering peak and low de-channelling rates (relatively low backscattering yield below 1500 keV), suggesting that randomly displaced atoms dominate the channelling phenomena. In contrast, the spectra for $[\overline{10}\,0\,23]$ are dominated by de-channelling, suggesting the distortion of this channel by extended defects. When only measuring one axis per sample, it is not clear whether such discrepancy is due to the formation of different types of defects in samples with different surface orientation, or if the same defect causes different signatures when measuring along different axes due to shadowing by the matrix atoms.

In order to further understand how the channelling phenomenon is affected by measuring different crystallographic axes, for the same surface orientation $(\bar{2}01)$, RBS/C spectra were measured along the [001], $[\bar{1}00]$, $[\bar{2}05]$ and $[\bar{10}\,0\,23]$ channels, (in addition to the random spectrum), as shown in Fig. 6. Just like above, the respective DECO relative defect concentration profiles for each fluence were calculated as well. Moreover, the corresponding $\Delta\chi_{\min}$ values are also shown in Fig. 5.

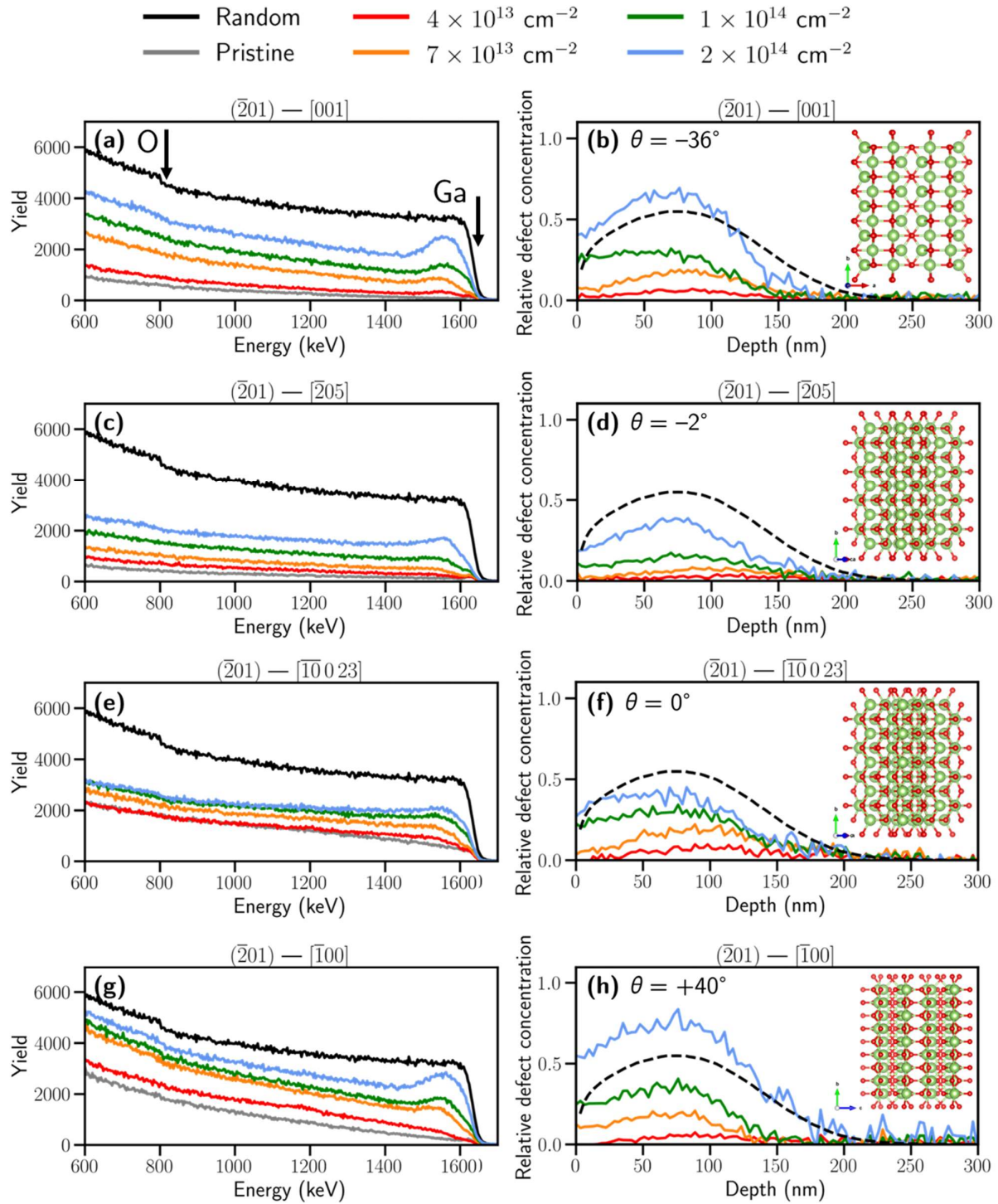

**Fig. 6** | RBS/C analysis for the $(\bar{2}01)$-oriented samples along the [001] (a–b), $[\bar{2}05]$ (c–d), $[\overline{10}\,0\,23]$ (e–f) and $[\bar{1}00]$ channels (g–h), for increasing implantation fluence, as well as the corresponding random spectra for reference. Panels (a), (c), (e), (g) show the experimental spectra, while panels (b), (d), (f) and (h) show the relative defect concentration profiles as calculated by DECO. The dashed lines represent the vacancy profiles obtained from SRIM Monte Carlo simulations, in an arbitrary scale. The insets show a sketch of the view of the crystal structure along the aligned axes, obtained using the VESTA software [33]. The tilt angle $\vartheta$ with respect to the surface normal is also indicated for each channel.

The different evolution of the spectral shapes and yields for each of the probed channels is very clear from Fig. 6. The $[\bar{1}00]$ and [001] channels get fairly close to the random level, while $[\bar{2}05]$ and $[\overline{10}\,0\,23]$ do not go to defect fractions beyond ~50%. In all cases, however, it is possible to notice a steady increase of the yield of the defect peak with the fluence. Moreover, the DECO profiles clearly show a Gaussian-like distribution of the defects in a range that is compatible with the vacancy distribution predicted by SRIM, and there is a very good agreement regarding the damaged depth across all the channels. It is also interesting to note that, while the defect profiles for two close-to-normal channels ($[\bar{2}05]$ and $[\overline{10}\,0\,23]$) have similar maxima, the profiles for the channels at higher angles have significantly higher values irrespective of the channelling quality of the axes in the pristine condition (see Fig. 3). This difference suggests that some point defects may not be visible along the channels (almost) perpendicular to the surface, but may be visible when probing the tilted channels. On the other hand, the shape of the defects peak along the $[\bar{1}00]$ and [001] axes are similar and more Gaussian-like than that of the $[\bar{2}05]$ and $[\overline{10}\,0\,23]$, which may suggest a different nature of the defects observed along each direction.

In this context, there are multiple literature reports on RBS/C analysis of implantation damage along different crystallographic directions, but these are often limited to more symmetric systems, such as cubic Si. For example, when implanting (111)-oriented Si crystals with Sb, D. Sigurd and K. Björkqvist report minimum yields of 6.3, 5.2 and 14% when measuring along the ⟨111⟩, ⟨110⟩ and ⟨100⟩ channels, respectively [41]. Moreover, W. J. Weber et al. report a different evolution of the relative interstitial disorder with the fluence in Au-implanted SiC when the beam is channelled along the ⟨0001⟩ or ⟨10$\bar{1}$1⟩ axes. Specifically, this was attributed to the shadowing of the Si and C interstitials along the former channel [42]. Therefore, all these previous works show how the apparent defect concentration measured via RBS/C along different axes can be different even for the same sample, how the de-channelling properties can be different along each of the probed channels, and even how the detection of some defects may be hindered or enhanced by the choice of the appropriate axes. This anisotropic visibility of implantation-induced defects must therefore be carefully considered when assessing defect formation or recovery processes.

## 3.3. Damage recovery under thermal annealing

In order to probe the recovery of the implantation-induced damage, the sample implanted with a fluence of $2\times10^{14}$ cm$^{-2}$ was then subjected to RTA at temperatures up to 1000 °C for each surface orientations. Fig. 7 shows the RBS/C analysis performed as a function of the annealing temperature for the channels (almost) perpendicular to the surface planes. As before, the $\Delta\chi_{min}$ values were also calculated for each orientation and are shown in the right side of Fig. 5.

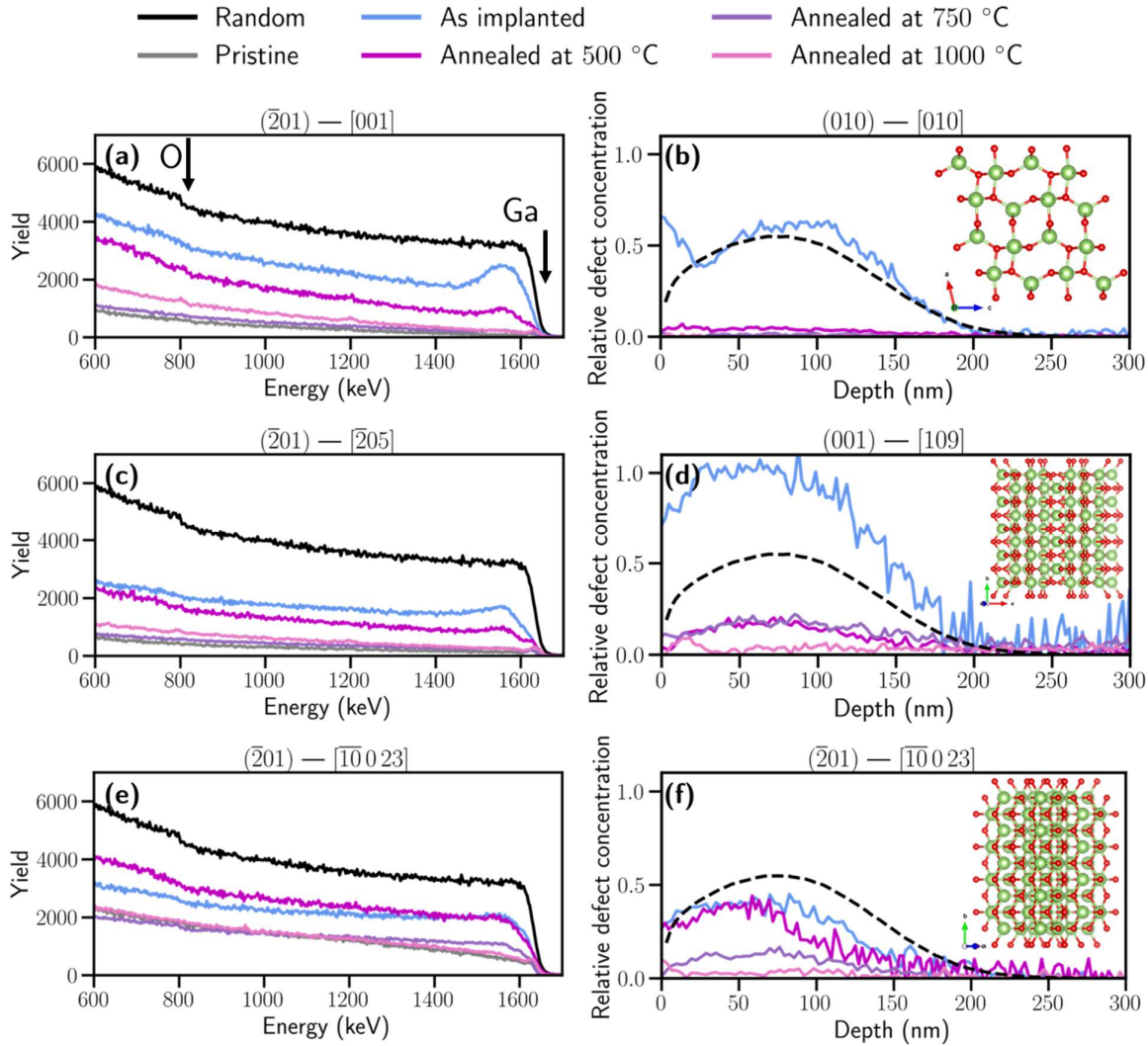

**Fig. 7** | RBS/C analysis for the (010)-oriented samples (a–b), along the [010] channel; (001)-oriented samples (c–d), along the [109] channel; and ($\bar{2}$01)-oriented samples (e–f), along the [$\overline{10}$ 0 23] channel, after implantation with a fluence of $2\times10^{14}$ cm$^{-2}$, for different annealing temperatures, as well as a random and a pristine spectrum for reference. Panels (a), (c) and (e) show the experimental spectra, while panels (b), (d) and (f) show the relative defect concentration profiles as calculated by DECO. The dashed lines represent the vacancy profiles obtained from SRIM Monte Carlo simulations, in an arbitrary scale. The insets show a sketch of the view of the crystal structure along the aligned axes, obtained using the VESTA software [33].

Regarding the (010)-oriented samples, we observe that annealing at a temperature as low as 500 °C already greatly suppresses the direct backscattering peak. It is important to mention that this temperature is close to the temperature needed to promote the transition from the $\gamma$-phase clusters that may be present back into the $\beta$-phase [43], as well as the temperature required to unroll $\beta$-Ga$_2$O$_3$ microtubes produced by ion implantation [10] or defects produced by proton irradiation [44,45]. This observation further supports the association between the direct backscattering and point defects, while extended defects require higher temperatures to be removed.

After annealing at 1000 °C, this peak essentially vanishes and the spectrum is almost perfectly superimposed with that of the pristine sample. This efficient defect removal is also verified in the extracted defect profiles, which essentially vanish for temperatures above 750 °C. In the case of the (001)-oriented sample, the decrease of the backscattering yield under annealing is not as efficient as for the previous case and progresses non-monotonically with the temperature. In particular, while the yield is greatly reduced after RTA at 500 °C, little spectral evolution is observed between this temperature and 750 °C. In fact, the two spectra are quite similar in the region of the defect peak, while there is a higher de-channelling for the higher temperature. Considering the way the de-channelling probability is accounted for in the context of the two-beam approximation, the DECO profiles end up being very similar for these two temperatures, suggesting a similar distribution of randomly displaced atoms but some changes in the de-channelling due to extended defects. After annealing at 1000 °C, the defect concentration profile is greatly reduced, but the spectrum never quite completely returns to the levels of the pristine sample. For the $(\overline{2}01)$-oriented sample measured along the $[\overline{10}\,0\,23]$ axis, annealing at 500 °C does not seem to have a big effect in the defect peak, but annealing at 750 and 1000 °C shows a clearer evolution with the temperature, leading to a good recovery at the latter temperature. All in all, these differences highlight how the complex microscopic processes associated with defect recovery can be assessed differently along different axes.

For the $(\overline{2}01)$-oriented sample, the RBS/C spectra were also measured along different channels, as shown in Fig. 8. The corresponding $\Delta\chi_{\text{min}}$ values were also calculated for each crystallographic axis and are shown in Fig. 5. With the exception of the $[\overline{10}\,0\,23]$ axis, annealing at 500 °C is already quite efficient in reducing the yield, which is attributed to the removal of point defects. However, for higher annealing temperatures, there are interesting differences in the spectra obtained along different channels. For the [001] channel, annealing at 750 °C promotes a clear reduction of the defect peak, as well as a reduction of the de-channelling, with the spectrum yield achieving values close to the pristine situation. However, this peak does not change much after annealing at 1000 °C, which is reflected by the near-superposition between the two DECO profiles, while the de-channelling contribution at lower energies increases at this temperature. On the other hand, we observe that, after annealing, the [001] and $[\overline{2}05]$ channels (i.e., the ones with the lowest pristine minimum yields) recover almost completely already at 750 °C, while the $[\overline{10}\,0\,23]$ and $[\overline{1}00]$ axes require a thermal treatment at 1000 °C. In any case, the damage recovery is quite remarkable for all the probed surface orientations and channels.

It is interesting to note that, in general, the recovery of the crystal at 500 °C seems to start in depth and proceeds towards the surface for all axes, while the de-channelling rates seem to increase. Since the de-channelling probabilities are often related with the presence of extended defects, this observation suggests that the annealing is efficient in removing the point defects that contribute to the direct backscattering peak in the RBS/C spectra, but extended defects remain and their concentration and size may even increase during annealing. Moreover, it is important to note that the spectra obtained for channels almost perpendicular to the surface, $[\overline{10}\ 0\ 23]$ and $[\overline{2}05]$, are dominated by de-channelling, suggesting that the induced point defects are shadowed in the direction perpendicular to the surface. This is in agreement with the fact that the 500 °C annealing appears to have little effect along these directions.

This annealing study was not performed for the (100)-oriented samples due to the removal of the implanted layer at a lower fluence. However, a previous study showed a very efficient defect removal upon RTA at 250 °C on Cr-implanted samples with a fluence of $5\times10^{13}$ cm$^{-2}$ [10], which corresponds to a dpa level 4 times lower than the one probed in the current work. On the other hand, after high-dpa (~40 dpa) implantation the temperature required for a full lattice recovery was about 1110–1200 °C for (100)-oriented $Ga_2O_3$ [12]. Therefore, this shows that the temperature required for annealing depends on the damage level. Moreover, the dependence of the point defect concentration on the annealing temperature has been previously reported to be non-monotonous, which is similar to what is observed in this work for the $[\overline{2}05]$ channel.

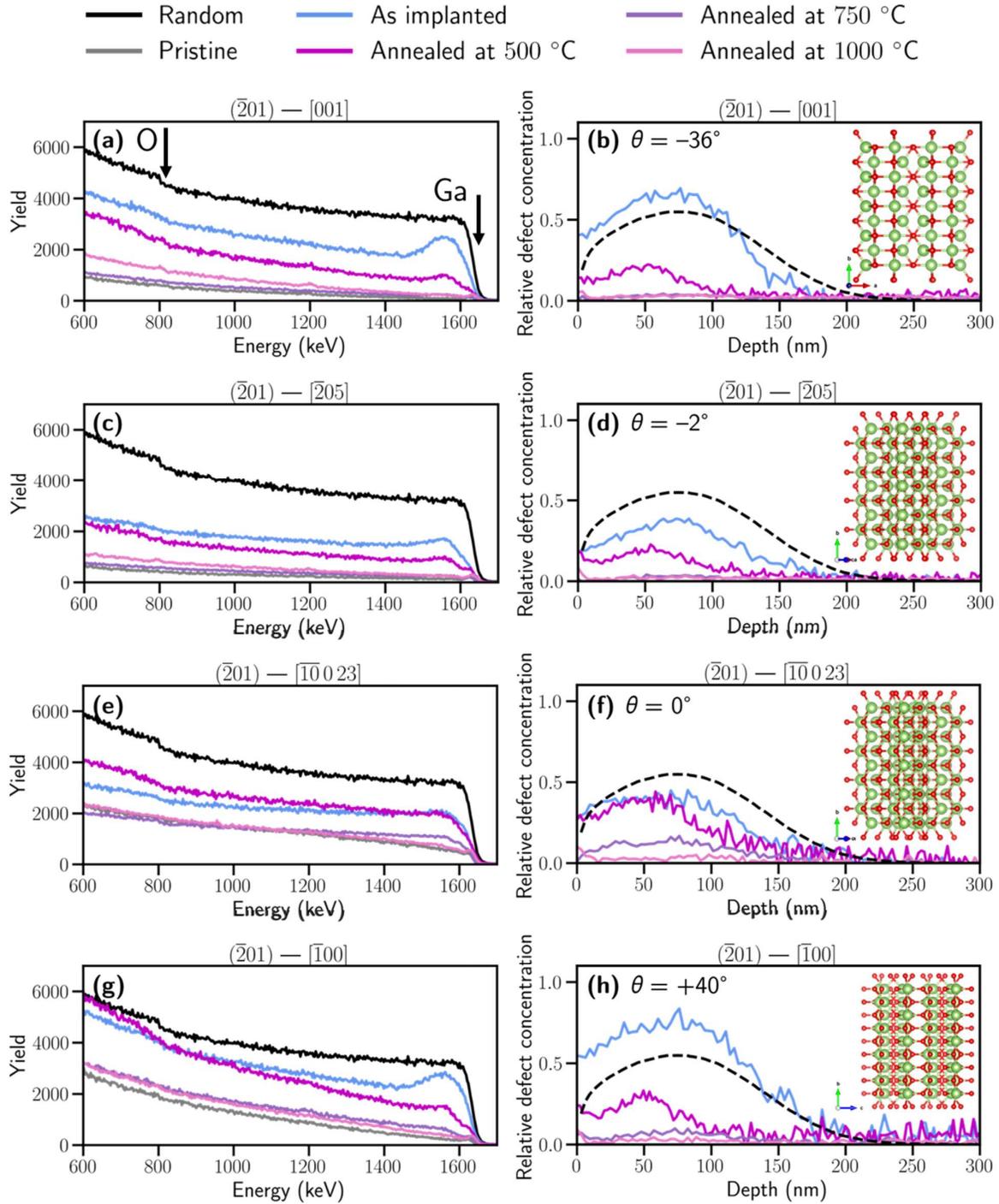

**Fig. 8** | RBS/C analysis for the $(\bar{2}01)$-oriented samples along the $[001]$ (a–b), $[\bar{2}05]$ (c–d), $[\overline{10}\,0\,23]$ (e–f) and $[\bar{1}00]$ channels (g–h), after implantation with a fluence of $2\times10^{14}$ cm$^{-2}$, for increasing annealing temperature, as well as the corresponding random spectra for reference. Panels (a), (c), (e) and (g) show the experimental spectra, while panels (b), (d), (f) and (h) show the relative defect concentration profiles as calculated by DECO. The dashed lines represent the vacancy profiles obtained from SRIM Monte Carlo simulations, in an arbitrary scale. The insets show a sketch of the view of the crystal structure along the aligned axes, obtained using the VESTA software [33]. The tilt angle $\vartheta$ with respect to the surface normal is also shown for each channel.

### 3.4. Strain relaxation via HRXRD

The efficiency of the thermal annealing processes was also assessed by HRXRD, by measuring its impact on the strain relaxation. The strain and the crystalline quality (quantified by the Debye-Waller factor, DW) profiles extracted for the (010)-, (010)- and (001)-oriented samples by HRXRD measurements for different implantation fluences have been published elsewhere [10,46]. In short, the previous results revealed no in-plane strain, and the accumulation of strain occurred in the out-of-plane direction only. Notably, the signs of the strains were different along each surface normal. Specifically, the strain is negative (compressive) along the direction perpendicular to the (010) plane, but positive (tensile) along the direction perpendicular to the (001) plane, in agreement with previous reports [19]. The opposite signs of the induced perpendicular strain along different surface normals further confirm the strong elastic anisotropy of $\beta$-$Ga_2O_3$ under ion implantation. For both directions, a steady decrease of DW was accompanied by an increase of the absolute value of the strain with increasing implantation fluence.

The out-of-plane strain accumulation was also assessed for the $(\bar{2}01)$ surface orientation with HRXRD via $2\vartheta$–$\omega$ scans about the $20\bar{1}$ symmetric reflection, as shown in Fig. 9 (a). For this sample, we observe that the pristine sample only shows a well-defined peak, while the implanted samples exhibit a shoulder for lower $2\theta$ values. This suggests an increase of the interplanar distance, i.e., the perpendicular strain $\varepsilon_\perp$ is tensile, in agreement with previous studies [16]. The maximum $\varepsilon_\perp$ increases up to about 1.4% at the highest implantation fluence. Note that no MROX fits were performed for this orientation since the strain accumulation occurs along both the $a$ and $c$ axes. Thus, it is not possible to resolve the strain along each axis directly from a single symmetric measurement.

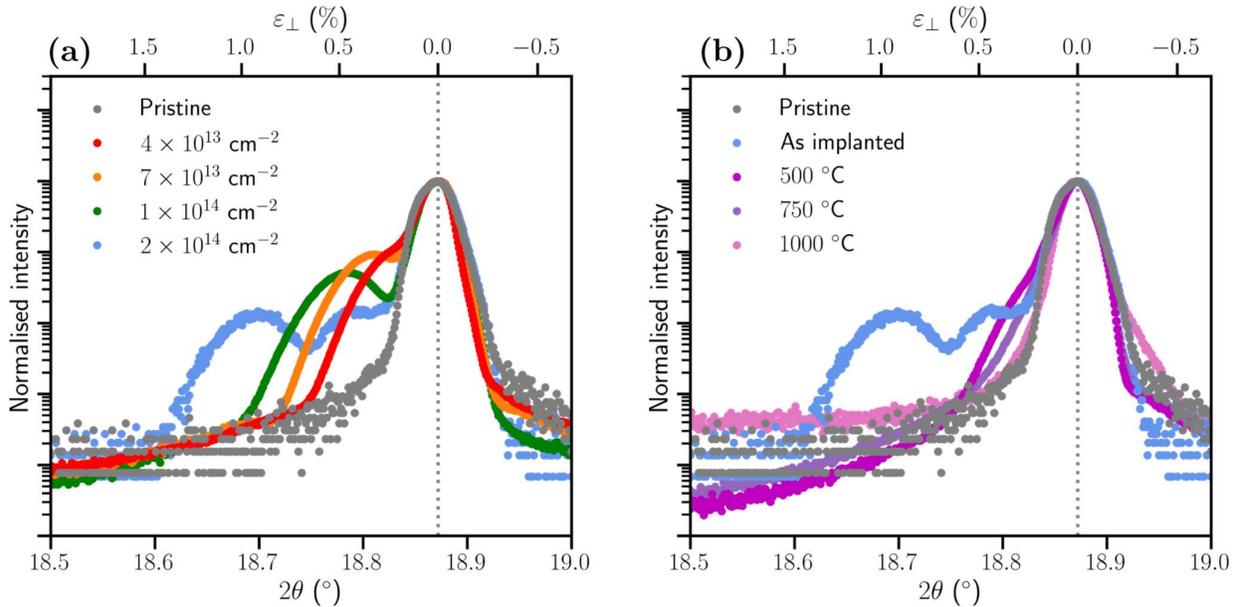

**Fig. 9** | HRXRD $2\vartheta$–$\omega$ scans about the $20\bar{1}$ symmetric reflection for the $(\bar{2}01)$-oriented samples, as a function of the implantation fluence (a) or the annealing temperature, after implantation with a fluence of $2\times10^{14}$ cm$^{-2}$ (b).

The experimental 2ϑ–ω scans about the 020 and 004 reflections, respectively, for the (010)- and (001)-oriented samples implanted to the highest fluence of $2\times10^{14}$ cm$^{-2}$ and subsequently annealed are shown in Fig. 10, as well as the evolution of the perpendicular strain, $\varepsilon_\perp$, and DW profiles obtained via MROX fits. Reciprocal space maps about selected reflections are shown in the supplementary information (Figs. S1 and S2), suggesting that, similarly to what was observed in the fluence dependence study [46], no measurable in-plane strain is present during the annealing processes within the resolution of the RSM. Note also the conversion from the $2\theta$ axis to $\varepsilon_\perp$ via Bragg's law, according to $\varepsilon_\perp = (\sin\theta_0 - \sin\theta)/\sin\theta$, where $\theta_0$ is the diffraction angle corresponding to the pristine sample. The out-of-plane strain recovery with annealing was also assessed for the $(\bar{2}01)$-oriented samples by HRXRD 2ϑ–ω scans about the $20\bar{1}$ symmetric reflection, as shown in Fig. 9 (b).

In agreement with the efficient defect removal suggested by RBS/C, a very good recovery of the crystalline quality and strain relaxation are observed upon RTA. HRXRD suggests that a large portion of the accumulated strain is already recovered at 500 °C, while RBS/C suggests an efficient removal of the point defects at this temperature. The combination of these observations suggests a possible correlation between the macroscopic strain state and the disorder associated with point-like defects detected by RBS/C.

It is also interesting to notice that the sign of the strain is reversed for the (010)-oriented sample, i.e., although the lattice parameter was first compressed (with respect to the pristine sample) after the ion implantation, it changed to tensile strain after annealing at 500 °C. Previous works report similar observations [14,47], which may be related to the removal of defects responsible for compressive strain at this temperature (likely vacancies), and to the formation/stabilisation of defect complexes introducing tensile strain. Regarding the (100) surface orientation, a previous study performed on a Cr-implanted sample up to $5\times10^{13}$ cm$^{-2}$ showed that the introduced tensile strain, which achieved a maximum value of about 1%, was efficiently removed via RTA, with a nearly-complete recovery being observed at 1000 °C [10].

It is also worth pointing out that, in the case of the $(\bar{2}01)$-oriented samples, the reduction of the out-of-plane strain after annealing at 500 °C, was not as pronounced as for the (001) surface orientation, which may hint at the formation of different defect types. Moreover, since the reduction of the defect concentration profiles was more evident along the channels at higher angles than along those almost perpendicular to the surface, this additionally suggests that the nature of the point defects responsible for the strains induced perpendicularly to the surface is such that they tend to be less visible along that direction via RBS/C measurements.

However, in all orientations after annealing at 1000 °C, perpendicular strain values of almost 0% and a DW factor of almost 1 were observed, showing the efficient lattice recovery. Overall, the combined RBS/C and HRXRD analyses reveal that both the magnitude and the sign of the implantation-induced strain, as well as its thermal recovery, are strongly dependent on the considered crystallographic orientation. While RBS/C primarily probes defects visible along specific channels, HRXRD captures the macroscopic lattice response, and the consistent trends observed between both techniques support the interpretation that point-defect-like disorder plays a dominant role in the strain build-up and relaxation processes. These results further highlight the importance of considering crystallographic anisotropy when analysing implantation damage and recovery in monoclinic $\beta$-Ga$_2$O$_3$.

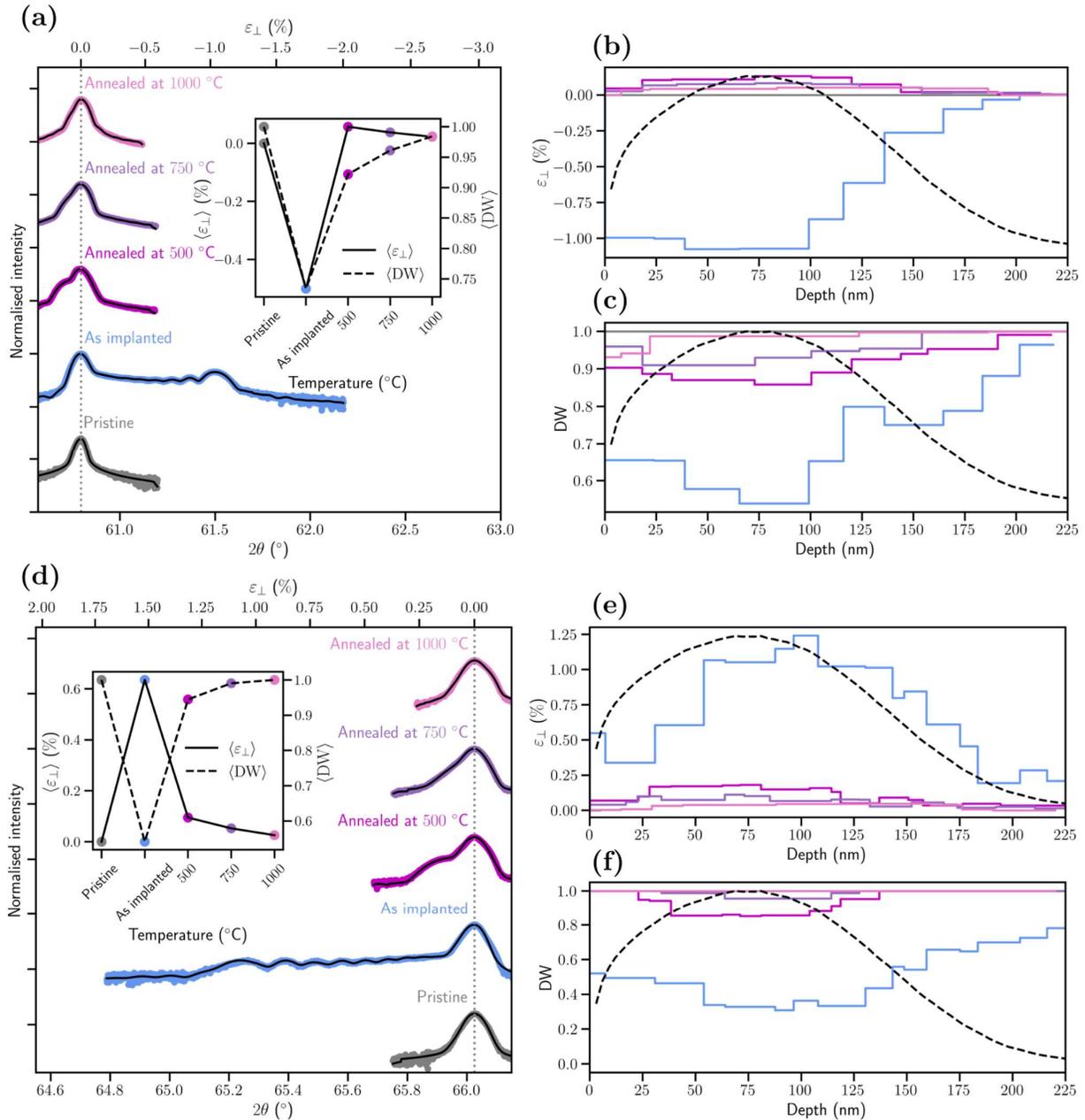

**Fig. 10** | HRXRD analysis for the (010)-oriented samples (a–c), about the 020 reflection, and (001)-oriented samples, (d–f), about the 004 reflection, after implantation with a fluence of $2\times10^{14}$ cm$^{-2}$, for different annealing temperatures. Panels (a) and (d) show the experimental diffractograms (points) and the MROX simulations (black lines), in a logarithmic scale. The diffractograms were shifted vertically for visual clarity. The corresponding perpendicular strain, $\varepsilon_\perp$, profiles are shown in panels (b) and (e), while panels (c) and (f) show those of the Debye-Waller factors, DW. The dashed lines represent the vacancy profiles obtained from SRIM Monte Carlo simulations, plotted in an arbitrary scale. The insets show the average values of $\varepsilon_\perp$ and DW as a function of the annealing temperature.

## 4. Conclusions

In this work, we assessed the channelling quality of multiple crystallographic axes in pristine $\beta$-Ga$_2$O$_3$, as well as the accumulation of defects under 250 keV Cr$^+$ implantations performed into samples with different surface orientations, namely (100), (010), (001) and $(\bar{2}01)$, as well as the removal of implantation-induced damage.

The RBS/C results show a clear structural anisotropy of the monoclinic system, as confirmed by the different minimum yields observed along the different channels already in the pristine sample. Different crystallographic axes were measured for different surface orientations, exhibiting minimum yield values ranging between 2 and 19%. Specifically, the hierarchy in terms of increasing minimum yield (decreasing channelling quality) was:

$$[010] \rightarrow [001] \rightarrow [\bar{2}05] \rightarrow [201] \rightarrow [\bar{1}00] \rightarrow [109] \rightarrow [\overline{10}\,0\,23].$$

After ion implantation, differences were also observed in terms of defect accumulation rates as a function of the fluence considering both the channels perpendicular to each surface plane and different channels in the $(\bar{2}01)$-oriented sample. Some of these axes, such as [010], showed high defect accumulation rates and pronounced direct backscattering peaks, which were attributed to point defects, small clusters of defects, or amorphous or $\gamma$-phase inclusions. Other axes, such as $[\overline{10}\,0\,23]$, show a low direct backscattering and high de-channelling, which has been attributed to extended defects, such as stacking faults or dislocation loops. The fact that spectra obtained along different channels within the same sample exhibit different behaviours strongly suggests that the same defects can have a different signature when observed along different crystallographic axes.

Remarkably, a very efficient defect removal was observed for all surface orientations, and for all the probed channels, being accompanied by an almost complete strain relaxation at 1000 °C, as measured by HRXRD. For intermediate temperatures, the strain relaxation and relative defect fraction progressed differently along different directions and for different surface orientations. The careful comparison between HRXRD and RBS/C suggested that the strain is mostly induced by point defects, which give rise to high direct backscattering in RBS/C. These are removed efficiently at relatively low temperatures, yielding a great decrease of $\Delta\chi_{min}$ along the [010] channel annealing at temperatures as low as 500 °C, accompanied by significant strain relaxation. In contrast, the almost-absent evolution of $\Delta\chi_{min}$ observed for the $[\overline{10}\,0\,23]$ axis at this temperature suggests that higher temperatures are required to remove extended defects.

In short, this work contributes to a better understanding of the structural properties of monoclinic $\beta$-Ga$_2$O$_3$, in particular in the context of ion implantation. As such, this work adds to the current efforts of providing the physical basis for the future usage of ion implantation into $\beta$-Ga$_2$O$_3$, leveraging on the comprehension of the damage accumulation and recovery phenomena taking place in single-crystals with different surface orientations, as well as along different crystallographic axes.

**Acknowledgements**

The authors acknowledge funding of the Research Unit INESC MN from the Fundação para a Ciência e a Tecnologia (FCT) through Plurianual financing (UIDB/05367/2025, UID/PRR/5367/2025 and UID/PRR2/05367/2025) (doi: https://doi.org/10.54499/UIS/PRR/05367/2025 and https://doi.org/10.54499/UID/PRR2/05367/2025), as well as via the IonProGO project (2022.05329.PTDC, http://doi.org/10.54499/2022.05329.PTDC). D. M. Esteves thanks FCT for his PhD grant (2022.09585.BD, https://doi.org/10.54499/2022.09585.BD). The implantations were performed under the scope of proposal 26001 of the ReMade@ARI project (https://doi.org/10.3030/101058414), funded by the European Union as part of the Horizon Europe call HORIZON-INFRA-2021-SERV-01 under grant agreement number 101058414 and co-funded by UK Research and Innovation (UKRI) under the UK government's Horizon Europe funding guarantee (grant number 10039728) and by the Swiss State Secretariat for Education, Research and Innovation (SERI) under contract number 22.00187. Neither the European Union nor any of the granting authorities can be held responsible for those opinions.

## 5. References

(1) Qin, Y.; Wang, Z.; Sasaki, K.; Ye, J.; Zhang, Y. Recent Progress of Ga2O3 Power Technology: Large-Area Devices, Packaging and Applications. *Jpn. J. Appl. Phys.* **2023**, *62* (SF), SF0801. https://doi.org/10.35848/1347-4065/acb3d3.

(2) Ping, L. K.; Berhanuddin, D. D.; Mondal, A. K.; Menon, P. S.; Mohamed, M. A. Properties and Perspectives of Ultrawide Bandgap $Ga_2O_3$ in Optoelectronic Applications. *Chin. J. Phys.* **2021**, *73*, 195–212. https://doi.org/10.1016/j.cjph.2021.06.015.

(3) Sasaki, K.; Higashiwaki, M.; Kuramata, A.; Masui, T.; Yamakoshi, S. MBE Grown $Ga_2O_3$ and Its Power Device Applications. *J. Cryst. Growth* **2013**, *378*, 591–595. https://doi.org/10.1016/j.jcrysgro.2013.02.015.

(4) Zheng, G.; Lou, C.; Yuan, Z.; Xiao, W.; Shang, L.; Zhong, J.; Tang, M.; Qiu, J. Rare-Metal-Free Ultrabroadband Near-Infrared Phosphors. *Adv. Mater.* **2025**, *37* (4), 2415791. https://doi.org/10.1002/adma.202415791.

(5) Zhang, Q.; Li, N.; Zhang, T.; Dong, D.; Yang, Y.; Wang, Y.; Dong, Z.; Shen, J.; Zhou, T.; Liang, Y.; Tang, W.; Wu, Z.; Zhang, Y.; Hao, J. Enhanced Gain and Detectivity of Unipolar Barrier Solar Blind Avalanche Photodetector via Lattice and Band Engineering. *Nat. Commun.* **2023**, *14* (1), 418. https://doi.org/10.1038/s41467-023-36117-8.

(6) Polyakov, A. Y.; Yakimov, E. B.; Shchemerov, I. V.; Vasilev, A. A.; Kochkova, A. I.; Nikolaev, V. I.; Pearton, S. J. Huge Photosensitivity Gain Combined with Long Photocurrent Decay Times in Various Polymorphs of $Ga_2O_3$: Effects of Carrier Trapping with Deep Centers. *J. Phys. Appl. Phys.* **2024**, *58* (6), 063002. https://doi.org/10.1088/1361-6463/ad8e6e.

(7) Hou, X.; Zhao, X.; Zhang, Y.; Zhang, Z.; Liu, Y.; Qin, Y.; Tan, P.; Chen, C.; Yu, S.; Ding, M.; Xu, G.; Hu, Q.; Long, S. High-Performance Harsh-Environment-Resistant $GaO_X$ Solar-Blind Photodetectors via Defect and Doping Engineering. *Adv. Mater.* **2022**, *34* (1), 2106923. https://doi.org/10.1002/adma.202106923.

(8) Current, M. I. Ion Implantation of Advanced Silicon Devices: Past, Present and Future. *Mater. Sci. Semicond. Process.* **2017**, *62*, 13–22. https://doi.org/10.1016/j.mssp.2016.10.045.

(9) Azarov, A.; Fernández, J. G.; Zhao, J.; Djurabekova, F.; He, H.; He, R.; Prytz, Ø.; Vines, L.; Bektas, U.; Chekhonin, P.; Klingner, N.; Hlawacek, G.; Kuznetsov, A. Universal Radiation Tolerant Semiconductor. *Nat. Commun.* **2023**, *14* (1), 4855. https://doi.org/10.1038/s41467-023-40588-0.

(10) Esteves, D. M.; He, R.; Bazioti, C.; Magalhães, S.; Sequeira, M. C.; Santos, L. F.; Azarov, A.; Kuznetsov, A.; Djurabekova, F.; Lorenz, K.; Peres, M. Microtubes and Nanomembranes by Ion-Beam-Induced Exfoliation of $\beta$-$Ga_2O_3$. arXiv:2501.13055 **2025**. https://doi.org/10.48550/arXiv.2501.13055.

(11) Peres, M.; Lorenz, K.; Alves, E.; Nogales, E.; Méndez, B.; Biquard, X.; Daudin, B.; Víllora, E. G.; Shimamura, K. Doping $\beta$-$Ga_2O_3$ with Europium: Influence of the Implantation and Annealing Temperature. *J. Phys. Appl. Phys.* **2017**, *50* (32), 325101. https://doi.org/10.1088/1361-6463/aa79dc.

(12) K. Lorenz; M. Peres; M. Felizardo; J. G. Correia; L. C. Alves; E. Alves; I. López; E. Nogales; B. Méndez; J. Piqueras; M. B. Barbosa; J. P. Araújo; J. N. Gonçalves; J. Rodrigues; L. Rino; T. Monteiro; E. G. Villora; K. Shimamura. Doping of $Ga_2O_3$ Bulk Crystals and NWs by Ion Implantation; 2014; Vol. 8987, p 89870M. https://doi.org/10.1117/12.2037627.

(13) Wendler, E.; Treiber, E.; Baldauf, J.; Wolf, S.; Ronning, C. High-Level Damage Saturation below Amorphisation in Ion Implanted $\beta$-$Ga_2O_3$. *18th Int. Conf. Radiat. Eff. Insul. REI-18 Dates 26th 31st Oct. 2015* **2016**, *379*, 85–90. https://doi.org/10.1016/j.nimb.2016.03.044.

(14) Gann, K. R.; Pieczulewski, N.; Gorsak, C. A.; Heinselman, K.; Asel, T. J.; Noesges, B. A.; Smith, K. T.; Dryden, D. M.; Xing, H. G.; Nair, H. P.; Muller, D. A.; Thompson, M. O. Silicon Implantation and Annealing in $\beta$-$Ga_2O_3$: Role of Ambient, Temperature, and Time. *J. Appl. Phys.* **2024**, *135* (1), 015302. https://doi.org/10.1063/5.0184946.

(15) Sarwar, M.; Ratajczak, R.; Ivanov, V.; Turek, M.; Heller, R.; Wachnicki, L.; Wozniak, W.; Guziewicz, E. Structural Defects and Luminescence in Sm-Implanted $\beta$-$Ga_2O_3$. *Phys. Status Solidi RRL – Rapid Res. Lett.* **2025**, *19* (11), 2400415. https://doi.org/10.1002/pssr.202400415.

(16) Bektas, U.; Liedke, M. O.; Liu, H.; Ganss, F.; Butterling, M.; Klingner, N.; Hübner, R.; Makkonen, I.; Wagner, A.; Hlawacek, G. Defect Analysis of the $\beta$– to $\gamma$–$Ga_2O_3$ Phase Transition. *Adv. Funct. Mater.* **2025**, e09688. https://doi.org/10.1002/adfm.202509688.

(17) Iancu, D.; Zarkadoula, E.; Leca, V.; Hotnog, A.; Zhang, Y.; Weber, W. J.; Velişa, G. Intrinsic Property of Defective $\beta$-$Ga_2O_3$ to Self-Heal under Ionizing Irradiation. *Scr. Mater.* **2025**, *268*, 116858. https://doi.org/10.1016/j.scriptamat.2025.116858.

(18) Demchenko, I. N.; Syryanyy, Y.; Shokri, A.; Melikhov, Y.; Domagała, J.; Minikayev, R.; Derkachova, A.; Munnik, F.; Kentsch, U.; Zając, M.; Reck, A.; Haufe, N.; Galazka, Z. Local Structure Modification around Si Atoms in Si-Implanted Monocrystalline $\beta$-$Ga_2O_3$ (100) under Heated Substrate Conditions. *Acta Mater.* **2025**, *292*, 121036. https://doi.org/10.1016/j.actamat.2025.121036.

(19) Matulewicz, J.; Ratajczak, R.; Sarwar, M.; Grzanka, E.; Ivanov, V.; Kalita, D.; Mieszczynski, C.; Jozwik, P.; Prucnal, S.; Kentsch, U.; Heller, R.; Guziewicz, E. Comprehensive Structural and Optical

Analysis of Differently Oriented Yb-Implanted $\beta$-Ga$_2$O$_3$. *Phys. Status Solidi RRL – Rapid Res. Lett.* **2026** *20*, e2500060. https://doi.org/10.1002/pssr.202500060.

(20) Ratajczak, R.; Sarwar, M.; Kalita, D.; Jozwik, P.; Mieszczynski, C.; Matulewicz, J.; Wilczopolska, M.; Wozniak, W.; Kentsch, U.; Heller, R.; Guziewicz, E. Anisotropy of Radiation-Induced Defects in Yb-Implanted $\beta$-Ga$_2$O$_3$. *Sci. Rep.* **2024**, *14* (1), 24800. https://doi.org/10.1038/s41598-024-75187-6.

(21) Sarwar, M.; Ratajczak, R.; Mieszczynski, C.; Wierzbicka, A.; Gieraltowska, S.; Heller, R.; Eisenwinder, S.; Wozniak, W.; Guziewicz, E. Defect Accumulation in $\beta$-Ga$_2$O$_3$ Implanted with Yb. *Acta Mater.* **2024**, *268*, 119760. https://doi.org/10.1016/j.actamat.2024.119760.

(22) Alves, E.; Lorenz, K.; Catarino, N.; Peres, M.; Dias, M.; Mateus, R.; Alves, L. C.; Corregidor, V.; Barradas, N. P.; Fonseca, M.; Cruz, J.; Jesus, A. An Insider View of the Portuguese Ion Beam Laboratory. *Eur Phys J Plus* **2021**, *136*. https://doi.org/10.1140/epjp/s13360-021-01629-z.

(23) Wendler, E.; Wesch, W.; Götz, G. Radiation Damage and Optical Properties of Ar$^+$-Implanted GaP. *J. Appl. Phys.* **1991**, *70* (1), 144–149. https://doi.org/10.1063/1.350302.

(24) Bøgh, E. Defect Studies in Crystals by Means of Channeling. *Can. J. Phys.* **1968**, *46* (6), 653–662. https://doi.org/10.1139/p68-081.

(25) Lorenz, K.; Wendler, E.; Redondo-Cubero, A.; Catarino, N.; Chauvat, M.-P.; Schwaiger, S.; Scholz, F.; Alves, E.; Ruterana, P. Implantation Damage Formation in *a*-, *c*- and *m*-Plane GaN. *Acta Mater.* **2017**, *123*, 177–187. https://doi.org/10.1016/j.actamat.2016.10.020.

(26) Götz, G.; Gärtner, K. *High Energy Ion Beam Analysis of Solids*; Walter de Gruyter GmbH, 1988.

(27) Jozwik, P.; Nowicki, L.; Ratajczak, R.; Stonert, A.; Mieszczynski, C.; Turos, A.; Morawiec, K.; Lorenz, K.; Alves, E. Monte Carlo Simulations of Ion Channeling in Crystals Containing Dislocations and Randomly Displaced Atoms. *J. Appl. Phys.* **2019**, *126* (19), 195107. https://doi.org/10.1063/1.5111619.

(28) Nowicki, L.; Jagielski, J.; Mieszczyński, C.; Skrobas, K.; Jóźwik, P.; Dorosh, O. McChasy2: New Monte Carlo RBS/C Simulation Code Designed for Use with Large Crystalline Structures. *Nucl. Instrum. Methods Phys. Res. Sect. B Beam Interact. Mater. At.* **2021**, *498*, 9–14. https://doi.org/10.1016/j.nimb.2021.04.004.

(29) Caçador, A.; Jóźwik, P.; Magalhães, S.; Marques, J. G.; Wendler, E.; Lorenz, K. Extracting Defect Profiles in Ion-Implanted GaN from Ion Channeling. *Mater. Sci. Semicond. Process.* **2023**, *166*, 107702. https://doi.org/10.1016/j.mssp.2023.107702.

(30) Ziegler, J. F.; Biersack, J. P.; Littmark, U. *SRIM: The Stopping and Range of Ions in Solids*; Pergamon, 1985.

(31) Tuttle, B. R.; Karom, N. J.; O'Hara, A.; Schrimpf, R. D.; Pantelides, S. T. Atomic-Displacement Threshold Energies and Defect Generation in Irradiated $\beta$-Ga$_2$O$_3$: A First-Principles Investigation. *J. Appl. Phys.* **2023**, *133* (1), 015703. https://doi.org/10.1063/5.0124285.

(32) Pearton, S. J.; Yang, J.; Cary, P. H.; Ren, F.; Kim, J.; Tadjer, M. J.; Mastro, M. A. A Review of Ga$_2$O$_3$ Materials, Processing, and Devices. *Appl Phys Rev* **2018**, *5*. https://doi.org/10.1063/1.5006941.

(33) Momma, K.; Izumi, F. VESTA 3 for Three-Dimensional Visualization of Crystal, Volumetric and Morphology Data. *J. Appl. Crystallogr.* **2011**, *44* (6), 1272–1276. https://doi.org/10.1107/S0021889811038970.

(34) Magalhães Esteves, D.; Lorenz, K.; Peres, M. Algebraic & Numerical Crystallography Notebook. *Zenodo* **2025**. https://doi.org/10.5281/zenodo.14639740.

(35) Dygo, A.; Turos, A. Surface Studies of $A^{III}B^{V}$ Compound Semiconductors by Ion Channeling. *Phys. Rev. B* **1989**, *40* (11), 7704–7713. https://doi.org/10.1103/PhysRevB.40.7704.

(36) Lenkeit, K. Resonance Dechanneling in Strained-Layer Superlattices in the Case of Axial-to-Planar Channeling Transition. *Nucl. Instrum. Methods Phys. Res. Sect. B Beam Interact. Mater. At.* **1992**, *67* (1), 180–184. https://doi.org/10.1016/0168-583X(92)95797-U.

(37) Azarov, A.; Bazioti, C.; Venkatachalapathy, V.; Vajeeston, P.; Monakhov, E.; Kuznetsov, A. Disorder-Induced Ordering in Gallium Oxide Polymorphs. *Phys. Rev. Lett.* **2022**, *128* (1), 015704. https://doi.org/10.1103/PhysRevLett.128.015704.

(38) Kjeldby, S. B.; Azarov, A.; Nguyen, P. D.; Venkatachalapathy, V.; Mikšová, R.; Macková, A.; Kuznetsov, A.; Prytz, Ø.; Vines, L. Radiation-Induced Defect Accumulation and Annealing in Si-Implanted Gallium Oxide. *J. Appl. Phys.* **2022**, *131* (12), 125701. https://doi.org/10.1063/5.0083858.

(39) Huang, H.-L.; Chae, C.; Johnson, J. M.; Senckowski, A.; Sharma, S.; Singisetti, U.; Wong, M. H.; Hwang, J. Atomic Scale Defect Formation and Phase Transformation in Si Implanted $\beta$-$Ga_2O_3$. *APL Mater.* **2023**, *11* (6), 061113. https://doi.org/10.1063/5.0134467.

(40) Chang, C. S.; Tanen, N.; Protasenko, V.; Asel, T. J.; Mou, S.; Xing, H. G.; Jena, D.; Muller, D. A. $\gamma$-Phase Inclusions as Common Structural Defects in Alloyed $\beta$-$(Al_xGa_{1-x})_2O_3$ and Doped $\beta$-$Ga_2O_3$ Films. *APL Mater.* **2021**, *9* (5), 051119. https://doi.org/10.1063/5.0038861.

(41) Sigurd, D.; Björkqvist, K. Channeling Studies of Ion-Implanted Silicon. *Radiat. Eff.* **1973**, *17* (3–4), 209–220. https://doi.org/10.1080/00337577308232617.

(42) Weber, W. J.; Gao, F.; Jiang, W.; Zhang, Y. Fundamental Nature of Ion–Solid Interactions in SiC. *Nucl. Instrum. Methods Phys. Res. Sect. B Beam Interact. Mater. At.* **2003**, *206*, 1–6. https://doi.org/10.1016/S0168-583X(03)00680-3.

(43) Playford, H. Y.; Hannon, A. C.; Barney, E. R.; Walton, R. I. Structures of Uncharacterised Polymorphs of Gallium Oxide from Total Neutron Diffraction. *Chem. – Eur. J.* **2013**, *19* (8), 2803–2813. https://doi.org/10.1002/chem.201203359.

(44) Esteves, D. M.; Rodrigues, A. L.; Alves, L. C.; Alves, E.; Dias, M. I.; Jia, Z.; Mu, W.; Lorenz, K.; Peres, M. Probing the $Cr^{3+}$ Luminescence Sensitization in $\beta$-$Ga_2O_3$ with Ion-Beam-Induced Luminescence and Thermoluminescence. *Sci. Rep.* **2023**, *13* (1), 4882. https://doi.org/10.1038/s41598-023-31824-0.

(45) Peres, M.; Esteves, D. M.; Teixeira, B. M. S.; Zanoni, J.; Alves, L. C.; Alves, E.; Santos, L. F.; Biquard, X.; Jia, Z.; Mu, W.; Rodrigues, J.; Sobolev, N. A.; Correia, M. R.; Monteiro, T.; Ben Sedrine, N.; Lorenz, K. Enhancing the Luminescence Yield of $Cr^{3+}$ in $\beta$-$Ga_2O_3$ by Proton Irradiation. *Appl Phys Lett* **2022**, *120*. https://doi.org/10.1063/5.0089541.

(46) Esteves, D. M.; He, R.; Magalhães, S.; Sequeira, M. C.; Costa, A. R. G.; Zanoni, J.; Rodrigues, J.; Monteiro, T.; Djurabekova, F.; Lorenz, K.; Peres, M. Understanding the Anisotropic Response of $\beta$-$Ga_2O_3$ to Ion Implantation arXiv:2603.06398 **2026**. https://doi.org/10.48550/arXiv.2603.06398.

(47) Azarov, A.; Galeckas, A.; Bektas, U.; Hlawacek, G.; Kuznetsov, A. Optical Absorption and Emission in Nitrogen-Implanted $Ga_2O_3$ Controlled by Dynamic Defect Annealing. *Adv. Opt. Mater.* **2025**, e03595. https://doi.org/10.1002/adom.202503595.

# Supporting Information

# Anisotropic implantation damage build-up and crystal recovery in $\beta$-Ga$_2$O$_3$

D. M. Esteves[1,2,*], S. Magalhães[2,3], Â. R. G. da Costa[4], K. Lorenz[1,2,3], M. Peres[1,2,3]

[1]     INESC Microsystems and Nanotechnology, Rua Alves Redol 9, Lisboa 1000-029, Portugal

[2]     Institute for Plasmas and Nuclear Fusion, Instituto Superior Técnico, University of Lisbon, Av. Rovisco Pais 1, Lisboa 1049-001, Portugal

[3]     Department of Nuclear Science and Engineering, Instituto Superior Técnico, University of Lisbon, Estrada Nacional 10, km 139.7, Bobadela 2695-066, Portugal

[4]     Centro de Ciências e Tecnologias Nucleares, University of Lisbon, Estrada Nacional 10, km 139.7, Bobadela 2695-066, Portugal

*     Corresponding author: duarte.esteves@tecnico.ulisboa.pt

*β-Ga$_2$O$_3$; Ion implantation; Rutherford backscattering spectrometry; Channelling*

## S1. Reciprocal space maps

Reciprocal space maps (RSM) of the (010)- and (001)-oriented samples were measured about selected reflections, both in symmetric and asymmetric geometry, in order to probe the strain along different out-of-plane and in-plane directions as a function of the annealing temperature for the sample implanted with a fluence of $2\times10^{14}$ cm$^{-2}$ with 250 keV Cr$^+$ ions.

Considering the relative intensity of the geometrically and kinematically accessible reflections, the 020, 110 and 022 reflections were measured for the (010)-oriented sample, while the 004, 204 and 024 reflections were chosen for the (001)-oriented sample. These are shown in Figs. S1 and S2 as a function of the annealing temperature. The RSM for the as implanted and pristine samples are also shown as a reference. In each figure, the perpendicular strain is calculated as $\varepsilon_\perp = (Q_{\perp 0} - Q_\perp)/Q_\perp$, where $Q_{\perp 0}$ and $Q_\perp$ are the perpendicular coordinates of the scattering vector $Q$ before and after implantation, respectively. This quantity can be read on the right-hand side axis.

Starting from the RSM for the as implanted sample, it is possible to observe a very intense peak that is associated with the pristine regions of the sample, as well as an additional structure that reflects the induced strain. For the (010)-oriented sample, this structure appears for larger $Q_\perp$ values, indicating a compression of the out-of-plane lattice constant, while for the (001)-oriented sample it appears for lower $Q_\perp$ values, suggesting an expansion of the out-of-plane lattice parameter. On the other hand, in both cases, no measurable in-plane strain is observed within the resolution of the RSM measurements, as no shift of the diffraction peak is detected along the parallel component of $\boldsymbol{Q}$, $Q_\parallel$ upon ion implantation.

The behaviour with increasing annealing temperature is also different for each surface orientation, and in agreement with the observations mentioned in the main text. Specifically, after annealing at 500 °C, the peaks associated with the implanted region of the (010)-oriented sample shift from larger to smaller $Q_\perp$ values, indicating a change from compressive to tensile strain. After this sign change, the strain is observed to decrease with the increasing temperature, finally achieving a peak similar to the pristine case. On the other hand, for the (001)-oriented samples, the strain remains tensile throughout, but is gradually reduced with increasing annealing temperature. In both cases, $Q_\parallel$ does not change, suggesting, that strain is predominantly accommodated along the out-of-plane direction. These reciprocal space maps further corroborate that implantation-induced strain in $\beta\text{-}Ga_2O_3$ is strongly orientation-dependent and primarily accommodated along the surface normal.

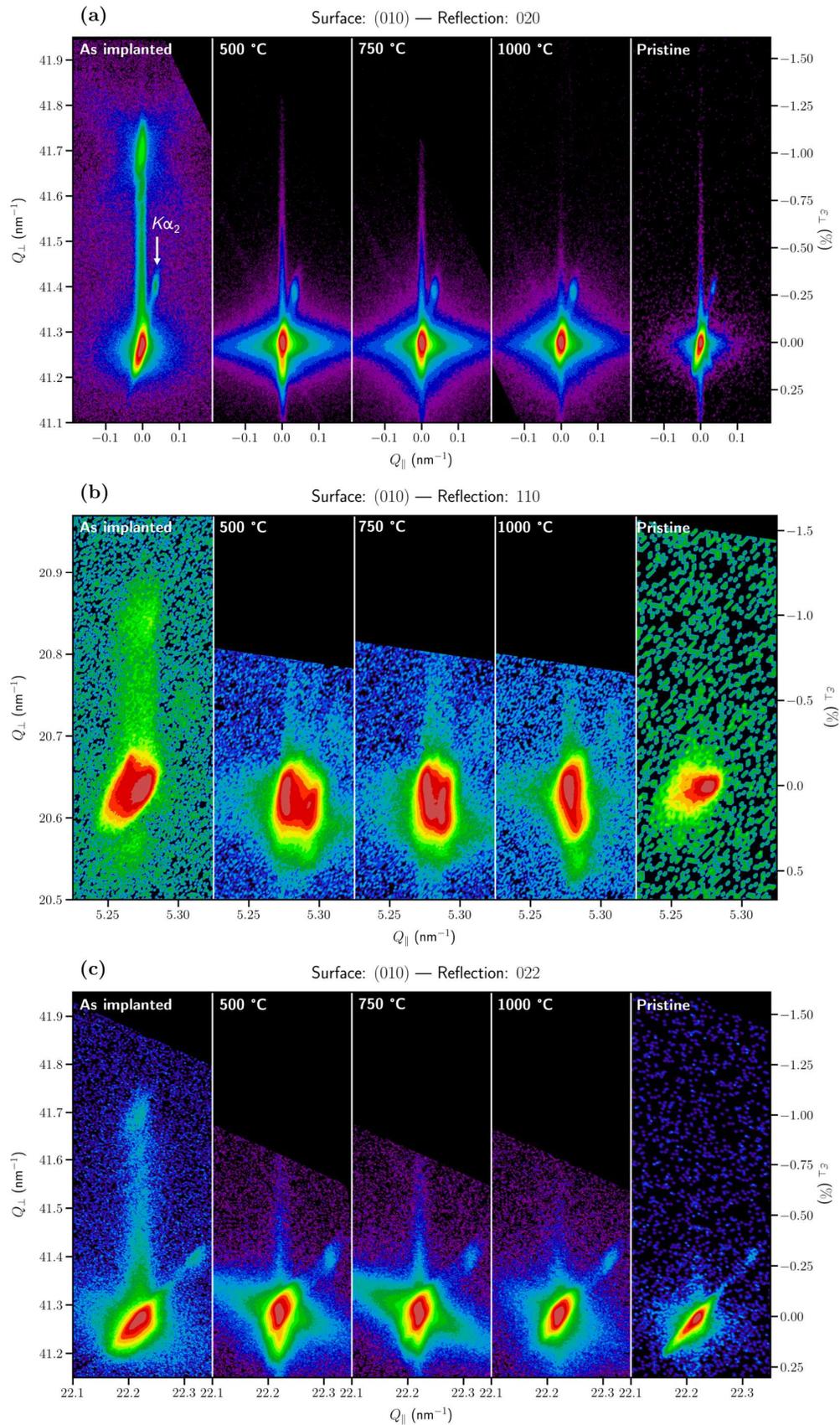

**Fig. S1** | Experimental RSM obtained for the (010)-oriented sample about the (a) 020, (b) 110 and (c) 022 reflections, as a function of the annealing temperature. The additional feature on the upper-right hand side of the most intense peak corresponds to a contribution due to the diffraction of the K$\alpha_2$ line.

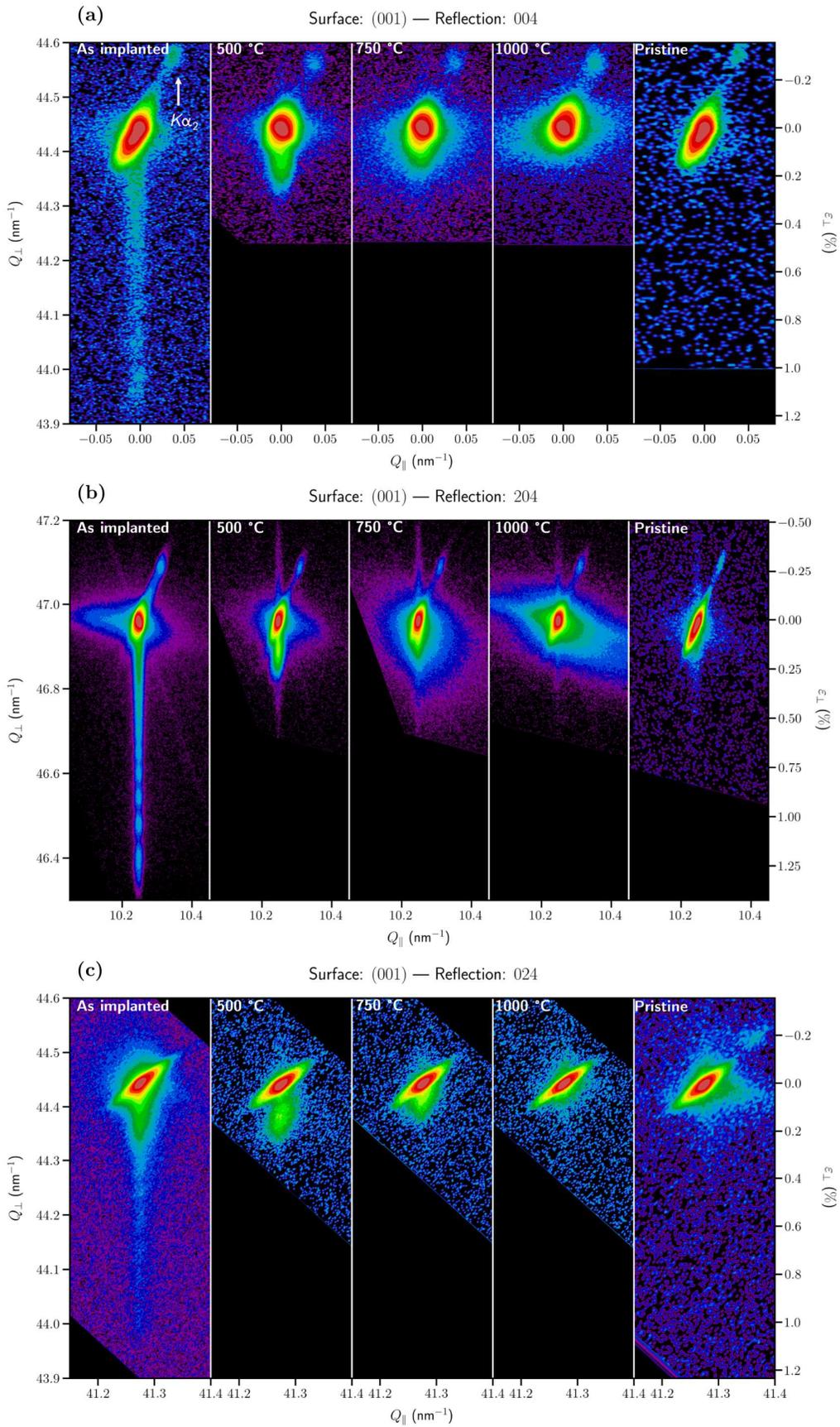

**Fig. S2** | Experimental RSM obtained for the (001)-oriented sample about the (a) 004, (b) 204 and (c) 024 reflections, as a function of the annealing temperature. The additional feature on the upper-right hand side of the most intense peak corresponds to a contribution due to the diffraction of the K$\alpha_2$ line.